\documentclass[preprint]{aastex}
\usepackage{epsf}
\newcommand{\simgt}{\lower.5ex\hbox{$\; \buildrel > \over \sim \;$}}
\newcommand{\simlt}{\lower.5ex\hbox{$\; \buildrel < \over \sim \;$}}
\newcommand{\cpara}{c_{\scriptscriptstyle \|}}
\newcommand{\cperp}{c_{\scriptscriptstyle \bot}}
\newcommand{\deltas}{\delta^{(s)}}
\newcommand{\sigmavar}{\sigma_{\scriptscriptstyle 0}}
\newcommand{\sigmapara}{\sigma_{\scriptscriptstyle 1,\|}}
\newcommand{\sigmaperp}{\sigma_{\scriptscriptstyle 1,\bot}}
\newcommand{\sigmatwo}{\sigma_{\scriptscriptstyle 2,\perp}}
\newcommand{\sigmaLN}{\sigma_{\rm \scriptscriptstyle LN}}
\newcommand{\PLN}{{\cal P}_{\rm \scriptscriptstyle LN}}
\newcommand{\LN}{{\rm \scriptscriptstyle LN}}
\newcommand{\sigmaP}{\sigma_{\rm \scriptscriptstyle P}}
\newcommand{\K}{\mathbf{k}}
\newcommand{\R}{\mathbf{r}}
\newcommand{\s}{\mathbf{s}}
\newcommand{\q}{\mathbf{q}}
\newcommand{\shb}{s_{\rm \scriptscriptstyle Hb}}
\newcommand{\seds}{s_{\rm \scriptscriptstyle Ed}}

\newcommand{\rs}{R_s}
\newcommand{\ks}{k_s}

\newcommand{\reff}{R_{\rm eff}}
\newcommand{\zmax}{z_{\rm \scriptscriptstyle max}}
\newcommand{\zmin}{z_{\rm \scriptscriptstyle min}}

\slugcomment{UTAP-374/2000 \quad HUPD-0007}
\shorttitle{Isodensity statistics of high-z objects}
\shortauthors{Taruya \& Yamamoto}
\begin{document}
\title{\Large\bf 
Isodensity Statistics on Clustering of High-$z$ Objects in 
Cosmological Redshift-spaces
}
%
%
%
%
%
\author{Atsushi Taruya$^1$ and Kazuhiro Yamamoto$^2$} 
\affil{ $^1$Department of Physics, University of Tokyo, Tokyo 113-0033, Japan \\
  $^2$Department of Physics, Hiroshima University, Higashi-Hiroshima 
  739-8526, Japan }
\email{ataruya@utap.phys.s.u-tokyo.ac.jp,  
kazuhiro@hiroshima-u.ac.jp}
%
%
%
%
%
%
%
%
\begin{abstract} 
We discuss the systematic effects arising from the
{\it cosmological redshift-space (geometric) distortion} 
on the statistical analysis of isodensity 
contour using high-redshift catalogs. 
Especially, we present a simple 
theoretical model for isodensity statistics in cosmological 
redshift-space, as a generalization of the work by Matsubara (1996). 
The statistical quantities considered here are the two- and 
three-dimensional genus of isodensity contour, 
the area of surface, the length of contour intersecting with a plane 
and the number of the crossing points of isodensity contour on a line. 
We give useful analytic formulae for the isodensity statistics, which 
take into account the corrections from the geometric distortion, the 
nonlinear clustering and the nonlinear velocity distortion 
phenomenologically. We then demonstrate how the geometric distortion 
and the nonlinear corrections 
alter shapes of the statistical quantities on the basis of plausible 
cosmological models. Our results show that the nonlinear correction 
can be sensitive to a choice of the redshift-space coordinate as 
increasing the redshift. 
The low-dimensional quantities such as 
two-dimensional genus systematically yield anisotropy 
due to the geometric and velocity distortions and their angle-dependence 
shows the $10\sim20\%$ difference of amplitude. 
Sensitivity for typical high-redshift samples are estimated in an 
analytic manner, and the influence of the light-cone effect for the 
isodensity statistics is also discussed. 
A simple estimation suggests that the systematic effects of 
geometric and redshift-space distortions 
can become comparable or could be dominant to the statistical error 
of deep cluster samples and future high-redshift galaxy surveys. 
These systematic effects might be a useful tool in 
probing the cosmological model of our universe.  
\end{abstract} 
%
%
%
\section{Introduction}
\label{sec: intro}
%
%
%
%
  Recent observations of high-redshift objects have revealed dynamical 
  aspects of the evolving universe. 
  Measurements of Lyman-break galaxies provide evidences for the early 
  galaxy assembly at the epoch $z\sim 3$, which put a serious constraint 
  on the structure formation scenario (Steidal et al. 1998; 
  Giavalisco et al. 1998). 
  Clusters up to $z\sim 1$ have been observed, and the abundance suggests 
  the universe with a low density parameter $\Omega \sim 0.3$ 
  (e.g, Carlberg et al. 1997). 
  Moreover, the upcoming wide-field surveys such as the Sloan Digital Sky 
  Survey (SDSS) and the Two-Degree Field (2dF) survey promise to extend the 
  observable scale of the universe and these enormous data will precisely 
  reveal the clustering feature of the high-redshift quasars and galaxies. 
  Since the analysis combining different redshift data sets has a potential 
  to break the degeneracy of the cosmological parameters, as well as to 
  provide the strong constraints on the structure formation scenario, 
  statistical studies using the high-redshift objects are of great 
  importance as a complementary 
  to the measurements of cosmic microwave background anisotropy and the 
  detection of cosmic shear.   

  Recently several authors have discussed various cosmological 
  effects arising from proper observations of the high-redshift 
  objects (e.g, Suto et al. 1999 and references therein). 
  Apart from the biasing of the luminous objects, the main 
  effects are summarized as follows:  

       \begin{enumerate}
       \item geometric distortion : no one can obtain the correct 
       three-dimensional map of the high-redshift objects without 
       knowing the correct cosmological model of the universe. The 
       uncertainty of the cosmological model may affect proper 
       evaluation of the clustering 
       pattern in the  {\it cosmological redshift space} 
       (Alcock \& Paczy\'nski 1979).
       \item redshift-space distortion :
       peculiar motion of a cosmological object affects the estimation
       of the distance in a redshift survey, which causes the linear 
       redshift-space distortion due to the bulk motion and the 
       nonlinear velocity distortion due to the virialized random motion 
       (finger-of-God effect) (Davis \& Peebles 1983; Kaiser 1987; 
       Suto \& Suginohara 1991; Hamilton 1998 for a review).  
       \item light-cone effect : because an entire sample does not consist 
       of objects on a constant-time hypersurface but on a light cone, the 
       redshift evolution of the objects inevitably contaminates the data. 
       Therefore, in a wide and deep survey, it becomes important to 
       incorporate the light-cone effect into statistical quantities, for 
       example, two-point correlation function and the power spectrum
       (Matarrese et al. 1997; Yamamoto \& Suto 1999). 
       \end{enumerate}
       
  For the popular two-point statistics, 
  the significance of these effects have been studied in detail 
  and the robustness of the theoretical prediction is checked by the 
  numerical simulation (Ballinger et al. 1996; Matsubara \& Suto 1996; 
  Magira et al. 2000; Hamana et al. 2000). 
  Using the theoretical formulae, the feasibility 
  of cosmological test with geometric distortion has been investigated 
  (Yamamoto, Nishioka \& Taruya 2000). Further, the analysis is developed 
  to the higher-order moments on a light-cone (Matsubara et al. 1997).

  Here we extend these analyses to the isodensity statistics. 
  That is, we investigate the influences of the
  above cosmological effects on the isodensity statistics, 
  which have not been ever clarified. 
  The isodensity statistics like the genus $G_3$ and area 
  statistic $N_3$ have progressively become 
  important as a quantitative measure of the topology 
  of the large-scale structure. Analytical expressions of 
  the isodensity statistics are derived for the Gaussian random 
  field and some cases of non-Gaussian field in real space (e.g, 
  Bardeen et al. 1986; Hamilton et al. 1986; Matsubara 1994, 2000; 
  Matsubara \& Yokoyama 1996). 
  With these analytic formulae, the isodensity statistics computed 
  from the redshift surveys of galaxies and/or the simulated dark matter 
  catalogues are comprehensively understood (Colley et al. 2000 
  and references therein).

  While most authors of these works have focused on the 
  low-redshift objects except for the cosmic microwave background 
  anisotropy, the analysis using the high-redshift samples is expected
  to provide an important clue on the origin and evolution of 
  proto-galaxies and proto-clusters. 
  In fact, Park et al. (2000) have recently analyzed the 
  two-dimensional genus using the galaxies taken from the Hubble Deep 
  Field and quantified the feature of early galaxy assembly. 
  Remarkably, their samples are widely distributed over the redshift range 
  $0\simlt z \simlt 3$, which revealed an unprecedentedly deep view of 
  galaxy clustering. Although the analysis is still limited to 
  the projected density field because of the large error of
  photometric redshift, precise measurement of redshift will enable 
  us to quantify the various isodensity statistics defined in 
  the three-dimensional space, as well as the quantities 
  in the one- and two-dimensional space in near future.

  Practically, the observed distribution of high-redshift objects 
  suffers from the cosmological effects, which should be correctly taken 
  into account when comparing such data with theoretical prediction. 
  Then naive but important questions are as follows: how the cosmological 
  effects affect the statistical study of isodensity contour; whether 
  these effects are observed as signal or noise. 
  Purpose of this paper is to address these issues by using a simple 
  theoretical formulae for the isodensity statistics, 
  in which the cosmological effects are faithfully incorporated. 
  Partially, the influence of the linear (redshift-space) distortion is 
  investigated by Matsubara (1996), 
  though the effect turns out to be small. In his paper the corrections  
  from the nonlinear clustering and the nonlinear velocity distortion, 
  which become influential on small length scales, are not investigated,  
  and neither is the geometric distortion.

  In this paper, we generalize the Matsubara's results to those 
  incorporating the geometric distortion effect and the nonlinear 
  correction taking account of the smoothing effect. With the 
  extended formulae, we examine how the geometric and redshift-space  
  distortions affect the isodensity statistics 
  qualitatively and quantitatively. We will show that the
  nonlinear effect can be important even for a case with large 
  smoothing scale because the scaling effect of the geometric 
  distortion causes apparent enhancement of the nonlinear effects, 
  which depends on choice of the cosmological redshift space. 
  In addition we also discuss the sampling noise and the effect of 
  finite redshift depth for typical high-redshift samples
  by using our extended formulae. 

  We organized the paper as follows.  In section 
  \ref{sec: Cosmological redshift space}, the cosmological redshift-space 
  coordinate is introduced, and the statistical quantities 
  incorporating the nonlinear corrections are described. 
  Then we consider the isodensity statistics in cosmological 
  redshift space in section \ref{sec: isodensity}. Based on the 
  analytic formulae for the Gaussian field in appendix 
  \ref{appendix: gaussian results}, we present a nonlinear model 
  of isodensity statistics. After demonstrating predictions of the  
  model, sensitivity of the isodensity statistics and the effect 
  of the light-cone are discussed in section \ref{sec: discussion}. 
  The final section \ref{sec: conclusion} is 
  devoted to summary and conclusion. 
%
%
\section{Density field in cosmological redshift space}
\label{sec: Cosmological redshift space}
%
%
%
%
\subsection{Redshift-space coordinate}
\label{subsec: coordinate}
%
%
%
%
%
We start with the standard form of Friedman-Robertson-Walker metric 
\begin{equation}
 ds^2=-c^2dt^2+a^2(t)\left[dr^2+f_K(r)^2
                      \left(d\theta^2+\sin^2{\theta}d\phi^2\right)\right],
\label{eq: metric}                    
\end{equation}
where $f_K(r)$ is the comoving angular diameter distance given by 
\begin{eqnarray}
\label{eq: comoving D_A}
 f_K(r)=\,\left\{
\begin{array}{lr}
\sin{(\sqrt{K}\,r)}/\sqrt{K} & (K>0) \\
r & (K=0) \\
\sinh{(\sqrt{-K}\,r)}/\sqrt{-K} & (K<0) \\
\end{array}
\right. , 
\end{eqnarray}
with the spatial curvature $K$ 
\begin{equation}
\label{eq: curvature}
 K\equiv \left(\frac{H_0}{c}\right)^2(\Omega_0+\lambda_0-1), 
\end{equation}
and the comoving radial distance 
\begin{equation}
\label{eq: comoving radial}
r(z)=\int_0^z\frac{c\,dz'}{H(z')}\,;~~~~~
H(z)=H_0\sqrt{\Omega_0(1+z)^3+(1-\Omega_0-\lambda_0)(1+z)^2+\lambda_0}.
\end{equation}
The quantities, $\Omega_0$, $\lambda_0$, and $H_0$, respectively, 
denote the density parameter, cosmological constant, and the 
Hubble parameter, at the present epoch $z=0$.

Because we are interested in a redshift survey, suppose that we could 
only know radial distances of objects in terms of the redshift $z$ 
and separations between pairs of objects in terms of the angular 
diameter distances. This means that the relation between the comoving 
real space and the observable redshift space inherently depends on 
an assumption of the cosmological model of the universe, and that 
the radial component of peculiar velocities of the objects apparently 
alters the distribution in the observed redshift space.

Hereafter, unless otherwise stated, we will assume that the sample of 
objects located in the range $z-dz$ and $z+dz$ with $dz\ll z$ whose 
angular separations are of the order of $dz$. Then, the 
distant-observer approximation is entirely valid. Within this 
approximation, we consider the object located at $\R=(r_1,r_2,r_3)$ 
in the comoving real space. 
{}For definiteness, 
we choose the third-axis as the line-of-sight direction.    
Denoting the corresponding coordinate in the observed cosmological 
redshift space by $\s=(s_1,s_2,s_3)$, 
the relation between $\s$ and $\R$ is given by 
\begin{eqnarray}
 s_{\scriptscriptstyle 1} &\equiv&
\frac{q_{\scriptscriptstyle 1}}{\cperp(z)}=\frac{r_{\scriptscriptstyle 1}}
{\cperp(z)},
\nonumber\\
 s_{\scriptscriptstyle 2} &\equiv& 
\frac{q_{\scriptscriptstyle 2}}{\cperp(z)}=\frac{r_{\scriptscriptstyle 2}}
{\cperp(z)},
\label{eq: s_vs_x}
\\
 s_{\scriptscriptstyle 3} &\equiv& 
\frac{q_{\scriptscriptstyle 3}}{\cpara(z)}=
\frac{1}{\cpara(z)}\left(r_{\scriptscriptstyle 3}+
\frac{1}{H(z)}\,v_{\scriptscriptstyle 3}\right),  
\nonumber
\end{eqnarray}
where $v_{\scriptscriptstyle 3}$ is the line-of-sight component of the 
peculiar velocity field. The observed redshift space $\s$ 
is obtained from the new coordinate system $\q=(q_{\scriptscriptstyle 1},q_{\scriptscriptstyle 2},q_{\scriptscriptstyle 3})$ 
divided by the dimensionless 
factors $\cperp$ and $\cpara$. Since we must assume a distance-redshift 
relation $s=s(z)$ to plot a map, the 
factors $\cperp$ and $\cpara$ 
depend on the assumption of the cosmological redshift space. 
{}Frequently, one adopts $s= \shb(z)$, the radial-distance 
relation extrapolating the Hubble law, and/or $s= \seds(z)$,
the relation in Einstein-de Sitter universe:
\begin{eqnarray}
        \shb(z) & \equiv &\frac{c\,z}{H_0}, 
\label{eq: shb}
\\
        \seds(z) & \equiv & \frac{2\,c}{H_0}
        \left(1-\frac{1}{\sqrt{1+z}}\right).
\label{eq: seds}
\end{eqnarray}
    Then, the geometric factors become 
\begin{equation}
\label{eq: z_crd}
 \cpara(z)\equiv\frac{d\,r(z)}{d\,s(z)}=\frac{H_0}{H(z)},
  ~~~~~~~~~~
 \cperp(z)\equiv\frac{f_K(r(z))}{s(z)}=\left(\frac{H_0}{c}\right)
 \frac{f_K(r(z))}{z},
\end{equation}
for the case $s=\shb(z)$ \cite{MS1996}, and    
\begin{equation}
\label{eq: EdS_crd}
 \cpara(z)=\frac{H_0}{H(z)}(1+z)^{3/2},~~~~~~~~~~
 \cperp(z)=\left(\frac{H_0}{c}\right)\frac{f_K(r(z))}{2(1-1/\sqrt{1+z})}, 
\end{equation}
for the case $s=\seds(z)$ \cite{BPH1996}, respectively. 
%
%
%
%
\subsection{Field correlations}
\label{subsec: field_corr}
%
%
%
%
Recall that the number density fluctuation of luminous objects 
is observed in the cosmological redshift space $\s$. We can evaluate 
the fluctuation $\deltas(\s)$ smoothed over the spherical radius $\rs$ : 
\begin{eqnarray}
 \deltas(\s;\rs)&=&
  \int d^3\s'\,\, W(\s-\s';\rs)\deltas(\s')
\nonumber \\
  &=&\int \frac{d^3\K_s}{(2\pi)^3}\hat{W}(\ks\rs)\hat{\delta}^{(s)}(\K_s)
  e^{-i\K_s\cdot\s}, 
\label{eq: smoothed_field}
\end{eqnarray}
where the quantities $\hat{W}$ and $\hat{\delta}^{(s)}$ are 
       the Fourier transform of the window function $W$ and the density 
       field $\deltas$, respectively. $\K_s$ is the Fourier vector 
       corresponding to the $\s$-space. In this paper, 
       we adopt the Gaussian window function, i.e., 
       $\hat{W}(y)=\exp(-y^2/2)$.
       The density fluctuations 
      defined in $\q$- and $\R$-space are merely related to 
      $\deltas$ through the transformation (\ref{eq: s_vs_x}). 

      For our purpose,  it is convenient to introduce 
      the new field variables:
\begin{equation}
 \label{eq: def_eta_zeta}
 \eta_{\scriptscriptstyle i} \equiv \partial_i \deltas, 
\,\,\,\,\,\,\,\,\,\,
 \zeta_{\scriptscriptstyle ij} \equiv 
\partial_{\scriptscriptstyle i}\partial_{\scriptscriptstyle j} 
\deltas \,\,; \,\,\,\,\,\,\,\,\,\,\,\, (i,j=1,2,3), 
\end{equation}
      where $\partial_{i}$ is the derivative with respect to the 
      cosmological redshift-space coordinate $s_{i}$. 
      Since correlation functions in the redshift space are 
      inherently anisotropic for the special direction of
      the line-of-sight, the variance and the cross-moments of the variables 
      (\ref{eq: def_eta_zeta}) have the following non-vanishing quantities: 
\begin{eqnarray}
\langle [\deltas]^2 \rangle &=& \sigmavar^2, 
\label{eq: var1} \\
\langle [\eta_{\scriptscriptstyle 3}]^2 \rangle &=& \sigmapara^2\, , 
\,\,\,\,\,
\langle \eta_{\scriptscriptstyle I}\eta_{\scriptscriptstyle J} 
 \rangle = \frac{\sigmaperp^2}{2} \delta_{\scriptscriptstyle IJ} 
\,\, =\,\, -\,\, \langle \zeta_{\scriptscriptstyle IJ}\deltas \rangle ,  
\label{eq: var2} \\
\langle \zeta_{\scriptscriptstyle IJ}\zeta_{\scriptscriptstyle KL} \rangle 
 &=& \frac{\sigmatwo^2}{24}
(\delta_{\scriptscriptstyle IJ}\delta_{\scriptscriptstyle KL}+
\delta_{\scriptscriptstyle IK}\delta_{\scriptscriptstyle JL}+
\delta_{\scriptscriptstyle IL}\delta_{\scriptscriptstyle JK}),
\label{eq: var3} 
\end{eqnarray}
where $\langle\cdots\rangle$ denotes ensemble average over the 
smoothed density fields. Subscripts $_I$ and $_J$ run from $1$ to $2$. 
The quantities $\sigmavar$, $\sigmapara$, $\sigmaperp$ and $\sigmatwo$ 
are related to the power spectrum $P^{\rm \scriptscriptstyle (crd)}$, 
defined in the cosmological redshift space $\s$. 
Substitution of (\ref{eq: smoothed_field}) 
into (\ref{eq: var1}), (\ref{eq: var2}) and (\ref{eq: var3}) yields 
\begin{eqnarray}
\label{eq: sigmavar}
 \sigmavar^2(\rs)&=&\int\frac{d^3\K_s}{(2\pi)^3}~\hat{W}^2(\ks\rs)
 ~ P^{\rm \scriptscriptstyle (crd)}
 (k_{\scriptscriptstyle s,\|},k_{\scriptscriptstyle s,\bot};z),
\\
\label{eq: sigmapara}
 \sigmapara^2(\rs)&=&
  \int\frac{d^3\K_s}{(2\pi)^3}~\hat{W}^2(\ks\,\rs)
  ~ P^{\rm \scriptscriptstyle (crd)}(k_{\scriptscriptstyle s,\|},k_{\scriptscriptstyle s,\bot};z)~
  k_{\scriptscriptstyle s,\|}^2, 
\\
\label{eq: sigmaperp}
 \sigmaperp^2(\rs)&=&
  \int\frac{d^3\K_s}{(2\pi)^3}~\hat{W}^2(\ks\,\rs)
 ~ P^{\rm \scriptscriptstyle (crd)}(k_{\scriptscriptstyle s,\|},k_{\scriptscriptstyle s,\bot};z)~
k_{\scriptscriptstyle s,\bot}^2,   
\\
\label{eq: sigmatwo}
 \sigmatwo^2(\rs)&=&
  \int\frac{d^3\K_s}{(2\pi)^3}~\hat{W}^2(\ks\,\rs)
 ~ P^{\rm \scriptscriptstyle (crd)}(k_{\scriptscriptstyle s,\|},k_{\scriptscriptstyle s,\bot};z)~
k_{\scriptscriptstyle s,\bot}^{4},   
\end{eqnarray}
with 
\begin{eqnarray}
  \ks=\sqrt{k_{\scriptscriptstyle s,\|}^2+k_{\scriptscriptstyle s,\bot}^2}, 
  \nonumber
\end{eqnarray}
where $k_{\scriptscriptstyle s,\|}(k_{\scriptscriptstyle 
s,\perp})$ denotes the wave number parallel(perpendicular) to the 
line-of-sight direction. 

We notice that the power spectrum 
$P^{\rm\scriptscriptstyle (crd)}(k_{\scriptscriptstyle s,\|},
k_{\scriptscriptstyle s,\bot};z)$ is written by
the power spectrum defined in the $\q$-space, 
$P^{(s)}(k_{\scriptscriptstyle \|},k_{\scriptscriptstyle \bot};z)$, 
where, $k_{\scriptscriptstyle \|}$ and $k_{\scriptscriptstyle \bot}$ are 
respectively the line-of-sight and its orthogonal component of the 
{}Fourier vector corresponding to the $\q$-space. Using the facts that 
$k_{\scriptscriptstyle s,\|}=\cpara k_{\scriptscriptstyle \|}$ 
and $k_{\scriptscriptstyle s,\bot}=\cperp k_{\scriptscriptstyle \bot}$, 
we obtain 
\begin{equation}
  \label{eq:Ps2Pcrd}
  P^{\rm\scriptscriptstyle (crd)}
        (k_{\scriptscriptstyle s,\|},k_{\scriptscriptstyle s,\bot}) 
   = \frac{1}{\cpara\,\cperp^{2}}\,\, 
 P^{(s)}\left(\frac{k_{\scriptscriptstyle s,\|}}{\cpara},
 \frac{k_{\scriptscriptstyle s,\bot}}{\cperp}\right).  
\end{equation}
Thus, all of the field correlations 
can be expressed in terms of the quantities defined in the $\q$-space:  
\begin{eqnarray}
\label{eq: sigmavar_P}
 \sigmavar^2(\rs)&=&\int\frac{d^3\K}{(2\pi)^3}~\hat{W}^2(k\,\reff)
 ~ P^{(s)}(k_{\scriptscriptstyle \|},k_{\scriptscriptstyle \bot};z),
\\
\label{eq: sigmapara_P}
 \sigmapara^2(\rs)&=&\cpara^2(z)
  \int\frac{d^3\K}{(2\pi)^3}~\hat{W}^2(k\,\reff)
  ~ P^{(s)}(k_{\scriptscriptstyle \|},k_{\scriptscriptstyle \bot};z)~
  k_{\scriptscriptstyle \|}^2, 
\\
\label{eq: sigmaperp_P}
 \sigmaperp^2(\rs)&=&\cperp^2(z)
  \int\frac{d^3\K}{(2\pi)^3}~\hat{W}^2(k\,\reff)
 ~ P^{(s)}(k_{\scriptscriptstyle \|},k_{\scriptscriptstyle \bot};z)~
k_{\scriptscriptstyle \bot}^2,   
\\
\label{eq: sigmatwo_P}
 \sigmatwo^2(\rs)&=&\cperp^4(z)
  \int\frac{d^3\K}{(2\pi)^3}~\hat{W}^2(k\,\reff)
 ~ P^{(s)}(k_{\scriptscriptstyle \|},k_{\scriptscriptstyle \bot};z)~
k_{\scriptscriptstyle \bot}^{4},   
\end{eqnarray}
where $k=\sqrt{k_{\scriptscriptstyle \|}^{2}+
  k_{\scriptscriptstyle \bot}^{2}}$. 
The quantity $\reff$ is related to the smoothing radius $\rs$ :  
\begin{equation}
\reff=\cperp\rs\sqrt{1-\left(\frac{k_{\scriptscriptstyle \|}}{k}\right)^{2}
\left\{1-\left(\frac{\cpara}{\cperp}\right)^{2}\right\}}.
\end{equation}
Note that the spherically symmetric smoothing in the $\s$-space 
corresponds to the anisotropic filtering in the $\q$-space in 
general cases with $\cpara\neq\cperp$. 
This can be regarded as an effect of the geometric distortion 
on the one-point statistics, as well as the isodensity statistics. 

Now we need a model for the power spectrum, 
$P^{(s)}(k_{\scriptscriptstyle \|},k_{\scriptscriptstyle \bot};z)$ 
(or $P^{\rm\scriptscriptstyle (crd)}$), 
including the redshift-space distortion and 
nonlinear clustering. 
Phenomenologically, but fairly accurate fitting form of the 
power spectrum has been recently proposed by several authors 
(Cole et al. 1995; Ballinger et al. 1996; Suto et al. 1999),
and its validity is successfully confirmed by N-body simulation 
(Magira et al. 2000). Hereafter, 
we assume that the observed number density fluctuation is simply 
proportional to the mass density contrast by introducing the 
biasing factor $b$ as the function of redshift   
\begin{equation}
  \label{eq: bias}
  \delta^{(r)}(\R,z) = b(z) \delta^{(r)}_{\rm mass}(\R,z), 
\end{equation}
in real space. Then the power spectrum of the observed 
density fluctuations is written as
\begin{equation}
  \label{eq: redshift-P(k)}
  P^{(s)}_{\rm nl}(k_{\scriptscriptstyle \|},k_{\scriptscriptstyle \bot};z) 
  = b^{2}(z) P^{(r)}_{\rm mass}(k)
  \,\,\left[1+\beta(z)\mu^{2}\right]^{2}\,\,D[k\mu\,\sigmaP(z)]\,\,  ,
\end{equation}
where $P^{(r)}_{\rm mass}(k)$ is the power spectrum of mass fluctuation 
in real space, $\sigmaP$ denotes the peculiar velocity dispersion, 
and we defined $\mu\equiv{k_{\scriptscriptstyle \|}}/{k}$. 
The factor $[1+\beta\,\mu^2]^2$ describes the linear 
theory of redshift-space distortion  (Kaiser 1987) and 
$\beta$ is obtained through the differentiation of the 
linear growth factor $D_+(z)$:  
\begin{equation}
  \label{eq: beta}
  \beta(z)= \frac{1}{b(z)}\,\,\frac{d \ln D_+(z)}{d \ln a} \simeq 
  \frac{1}{b(z)}\left[\Omega^{0.6}(z)+\frac{\lambda(z)}{70}
\left(1+\frac{\Omega(z)}{2}\right)\right],
\end{equation}
%
%
%
%
with 
\begin{equation}
  \label{eq: omega-lambda}
  \Omega(z)=\left(\frac{H_{0}}{H(z)}\right)^{2}(1+z)^{3}\,\Omega_{0}, 
  \,\,\,\,\,\,\,\,\,
  \lambda(z)=\left(\frac{H_{0}}{H(z)}\right)^{2}\,\lambda_{0}.
\end{equation}
The last term in (\ref{eq: redshift-P(k)}) is the damping factor 
due to the finger-of-God effect:
%
%
%
%
\begin{equation}
  \label{eq: damping_FOG}
  D[x] = \frac{1}{1+x^2/2},   
\end{equation}
which corresponds to the assumption of the exponential form for 
the distribution function of the pair-wise peculiar velocity in 
real space. To be specific, in this paper, we adopt the fitting 
formulae of Peacock \& Dodds (1996) and Mo et al. (1997) for the 
nonlinear mass power spectrum $P^{(r)}_{\rm mass}(k)$ and 
the peculiar velocity dispersion $\sigmaP$, 
respectively. Then the field correlations, 
$\sigmavar$, $\sigmapara$, $\sigmaperp$ and $\sigmatwo$ 
can be computed through the definitions 
(\ref{eq: sigmavar_P}-\ref{eq: sigmatwo_P}). 
%
%
%
%
%
\section{Isodensity statistics in cosmological redshift space}
\label{sec: isodensity}

In this section, we consider simple theoretical models
for isodensity statistics in cosmological redshift space. 
After describing definitions of the isodensity statistics, 
plausible nonlinear formulae are presented, based on 
the Gaussian results in appendix \ref{appendix: gaussian results}.  
Influences of the nonlinear corrections and the geometric distortion 
are quantitatively examined in familiar cold dark matter cosmological 
models.
%
%
\subsection{Definitions}
\label{subsec: definition}

{}First consider the three-dimensional genus curve. 
The genus $G_3$ is defined by $-1/2$ 
times the Euler characteristics 
of two-dimensional surfaces. The Euler number density is given by 
(number of holes) $-$ (number of isolated regions). Equivalently, 
this can be expressed by (number of maxima) $+$ (number of minima) 
$-$ (number of saddle points) of the isodensity contour with respect 
to some fixed direction irrespective of the isotropy of the random 
field. 
At a given density threshold $\nu\equiv\deltas/\sigmavar$, 
the genus $G_3$ is defined by (Bardeen et al. 1986)
\begin{equation}
\label{eq: def_Genus}
 G_3^{(s)}(\nu)=-\frac{1}{2}\,\,
\left\langle\delta_D(\deltas-\nu\sigmavar)
\delta_D(\eta_{\scriptscriptstyle 1})
\delta_D(\eta_{\scriptscriptstyle 2})
\,\,|\eta_{\scriptscriptstyle 3}|\,\,
\left(\zeta_{\scriptscriptstyle 11}\zeta_{\scriptscriptstyle 22}
  -\zeta_{\scriptscriptstyle 12}^2\right) \right\rangle,
\end{equation}
where $\delta_{D}(\cdot)$ denotes the Dirac's delta-function.

\begin{figure}
\begin{center}
   \leavevmode\epsfxsize=9cm \epsfbox{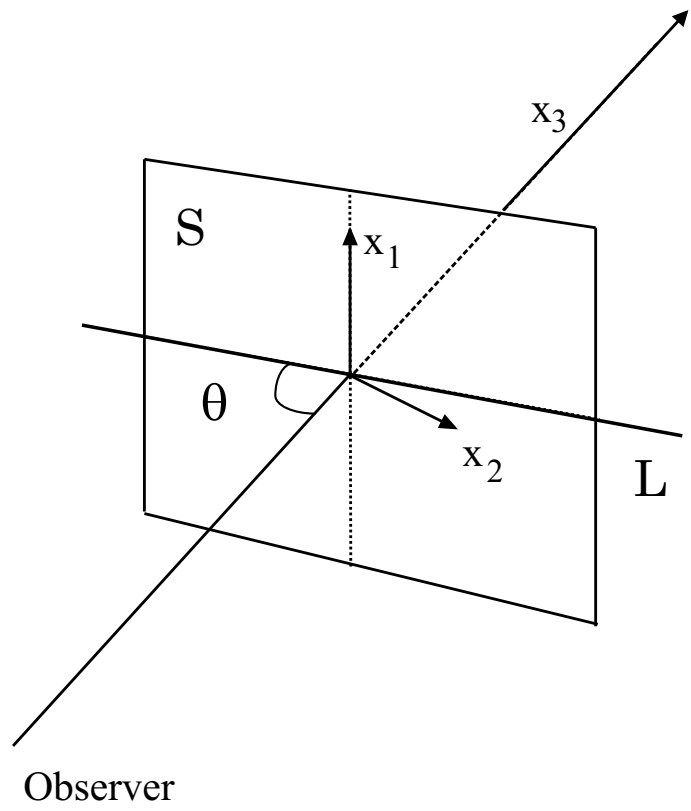}
\end{center}
\figcaption{ A sketch of cosmological redshift coordinate 
and the configuration of survey region (plane $S$/line $L$)
\label{fig: config}
}
\end{figure}

Genus is also defined for a two-dimensional random field. 
Consider the field in a two-dimensional flat plane $S$ 
and its isodensity contours. The two-dimensional genus $G_{2}$ 
can be computed by counting the number of the isolated 
high(low)-density regions surrounded by the isodensity 
contours (Melott et al. 1989). Notice that the quantity $G_{2}$ 
depends on a configuration of the plane $S$ because 
the distortion induced by the geometric and peculiar velocity
effects makes the random field anisotropy 
(Matsubara 1996). Denoting the angle 
between the line-of-sight direction and the plane $S$ by $\theta$ 
(see Fig.\ref{fig: config}), 
the two-dimensional genus per unit area, $G_{2}$, is 
\begin{equation}
\label{eq: def_2Dgenus}
 G_2^{(s)}(\nu,\theta)=-\frac{1}{2}\,\,
\left\langle\delta_D(\deltas-\nu\sigmavar)
\delta_D(\eta_{\scriptscriptstyle 1})
\,\,|\eta_{\scriptscriptstyle 2}\sin{\theta}
+\eta_{\scriptscriptstyle 3}\cos{\theta}|\,\,
\zeta_{\scriptscriptstyle 11} \right\rangle. 
\end{equation}

Next consider the area statistic $N_3$. This was proposed by 
Ryden (1988) and Ryden et al. (1989) as a complementary 
to the genus statistics. Given a density height $\nu$, 
the quantity $N_3$ quantifies the mean surface area of the 
isodensity contours per unit volume : 
\begin{equation}
\label{eq: def_N3}
 N_3^{(s)}(\nu)=\left\langle\delta_D(\deltas-\nu\sigmavar)
\sqrt{\eta_{\scriptscriptstyle 1}^2+\eta_{\scriptscriptstyle 2}^2+
\eta_{\scriptscriptstyle 3}^2}\,\right\rangle.
\end{equation}
{}For a homogeneous and isotropic random field, $N_3^{(s)}(\nu)$
equals $4/\pi$ times the mean length of the isodensity contour
with same height $\nu$ per unit area on a flat plane and twice 
the mean number of the crossing points per unit length on a straight 
line, which we denote by $N_2$ and $N_1$, respectively (Ryden 1988). 
In cosmological redshift space, these statistics obviously
depend on the angle $\theta$ between the line-of-sight direction 
and the plane $S$ or the line $L$ (Fig.\ref{fig: config}). 
Therefore we need treat $N_2$ and $N_1$, separately, which can 
be defined as
\begin{eqnarray}
\label{eq: def_N2}
 N_2^{(s)}(\nu,\theta) &=& \left\langle \delta_D(\deltas-\nu\sigmavar)
\sqrt{\eta_{\scriptscriptstyle 1}^2+
\left(\eta_{\scriptscriptstyle 2}\sin{\theta}+
    \eta_{\scriptscriptstyle 3}\cos{\theta}\right)^2}\,\right\rangle,  
\\
\label{eq: def_N1}
 N_1^{(s)}(\nu,\theta) &=& \left\langle \delta_D(\deltas-\nu\sigmavar)
\left|\eta_{\scriptscriptstyle 2}\sin{\theta}+\eta_{\scriptscriptstyle 3}
 \cos{\theta} \right|\,\right\rangle. 
\end{eqnarray}
%
%
%
%
%
%
\subsection{Modeling the nonlinear density field 
  and isodensity statistics}
\label{subsec: modeling}
%
%
%
%
Provided the distribution function for the field variables $\deltas$, 
$\eta_i$ and $\zeta_{ij}$, the expressions for the isodensity 
statistics defined in previous subsection can be derived analytically.  
{}For the homogeneous but anisotropic Gaussian random fields, 
we easily obtain the simple analytic formulae (appendix 
\ref{appendix: gaussian results}).  The influence of 
geometric distortion is naturally incorporated into them. 

The Gaussian results, however, are rather idealistic and cannot be 
used beyond the linear regime. Properly, we must take into 
account non-Gaussianity of the density distribution function developed 
by the gravitational clustering in the cosmological redshift space. 
Since no rigorous treatment of 
nonlinear evolution has been found, we need first adopt a reliable model of 
the distribution function ${\cal P}(\deltas,\eta_i,\zeta_{ij})$. 
A class of models for the nonlinear density field have been discussed, 
whose statistical property is characterized by a monotonic 
function $F$ (e.g, Coles \& Jones 1991): 
\begin{equation}
\label{eq: del-phi}
 \deltas(\s)=F[\phi(\s)], 
\end{equation}
where the scalar field $\phi(\s)$ is a Gaussian random variable with 
zero mean and the unit variance. 
In this case, it is easy to derive analytic expressions of 
isodensity statistics from the Gaussian results in appendix 
\ref{appendix: gaussian results}. 
This is partially done by Matsubara \& Yokoyama (1996) 
for the genus curve $G_3$. We generalize their result to   
the other statistics $G_2$, $N_3$, $N_2$ and $N_1$. 
In appendix \ref{appendix: nonGaussian}, 
assuming the functional form (\ref{eq: del-phi}), 
the analytic formulae for isodensity statistics 
in the non-Gaussian case 
are presented, which are simply obtained 
by transformation of the Gaussian results. 

The plausible functional form (\ref{eq: del-phi}) for the 
nonlinear density distribution has been investigated
by several numerical works. They showed that the probability 
density distribution of one-point statistics can be empirically 
fitted by the log-normal distribution function to a good 
accuracy (Kofman et al. 1994; Bernardeau \& Kofman 1995; 
Taylor \& Watts 2000). 
This is also the case in the redshift space 
(Hui et al. 2000). Observationally, the log-normal distribution 
function has been used as a model of the galaxy distribution 
ever since Hubble (1934) (e.g, Hamilton 1985; Bouchet et al. 1993; 
Kofman et al. 1994). Hence, we might use the log-normal function 
as a good approximation for the actual observed distribution.   

The log-normal distribution is characterized by   
\begin{equation}
  \label{eq: F(phi)}
  F(\phi) = \frac{1}{\sqrt{1+\sigmavar^{2}}}\,\,
  \exp\left[\sigmaLN\,\phi\right]\,-1,
\end{equation}
where
\begin{equation}
  \label{eq: sigma_LN}
  \sigmaLN^{2} = \ln (1+\sigmavar^{2}). 
\end{equation}
Then we obtain the one-point distribution function 
\begin{equation}
  \label{eq: lognormalPDF}
  \PLN(\deltas) d\deltas = \frac{1}{\sqrt{2\pi}
    \sigmaLN} 
  \exp\left[-\frac{\left\{\ln(1+\deltas)+
        \sigmaLN^{2}/2\right\}^{2}}
    {2\sigmaLN^{2}}\right]
  \frac{d\deltas}{1+\deltas}, 
\end{equation}
and 
\begin{equation}
\label{eq: relation_LN}
 F^{-1}(\nu\sigmavar)=\frac{\ln(1+\nu\sigmavar)+\sigmaLN^2/2}{\sigmaLN}, 
  \,\,\,\,\,\,\,\,\,\,\,\,
 \langle F_{,\phi}(\phi)^2\rangle_{\phi}=\sigmaLN^2(1+\sigmavar^2). 
\end{equation}
Using (\ref{eq: relation_LN}) and (\ref{C: G_phi})-(\ref{C:
N1_phi}), the analytic expressions for log-normal distribution are 
obtained as follows : 
\begin{eqnarray}
 G_{3,\LN}^{(s)}(\nu) 
\,&=& \frac{1}{4\pi^2(1+\sigmavar^2)^{3/2}}\,\,
\frac{\sigmaperp^2\sigmapara}{\sigmaLN^3}
\left[1-\left\{\frac{\ln(1+\nu\sigmavar)+\sigmaLN^2/2}{\sigmaLN}\right\}^2\right]
\nonumber\\
\label{eq: G_log} 
&&~~~~~~~~~~~~~~~~~~~~~~~~~~~ \times\,\,
 \exp\left\{-\frac{[\ln(1+\nu\sigmavar)+\sigmaLN^2/2]^2}
        {2\sigmaLN^2}\right\},
\end{eqnarray}
for the three-dimensional genus, and
\begin{eqnarray}
 G_{2,\LN}^{(s)}(\nu,\theta) 
\,&=&\, \frac{1}{8\pi(1+\sigmavar^2)}
  \left(\frac{\sigmaperp}{\sigmaLN}\right)^2
\,\,\frac{\ln(1+\nu\sigmavar)+\sigmaLN^2/2}{\sigmaLN} 
\nonumber \\
\label{eq: G2_log} 
&&~~~~~~~~~~~~~~~~~~~~~ \times\,\,
 \exp\left\{-\frac{[\ln(1+\nu\sigmavar)+\sigmaLN^2/2]^2}
        {2\sigmaLN^2}\right\}~L(\gamma,\theta),
\end{eqnarray}
for the two-dimensional genus. The quantities, $N_3$, $N_2$, and $N_1$  
reduce to
\begin{eqnarray}
\label{eq: N3_log}
 N_{3,\LN}^{(s)}(\nu) \,&=&\, \frac{1}{2\sqrt{\pi(1+\sigmavar^2)}}
  \frac{\sigmaperp}{\sigmaLN}
  ~\exp\left\{-\frac{[\ln(1+\nu\sigmavar)+\sigmaLN^2/2]^2}
        {2\sigmaLN^2}\right\}~~J(\gamma),
\\
\label{eq: N2_log}
 N_{2,\LN}^{(s)}(\nu,\theta) \,&=&\, \frac{1}{2\sqrt{\pi(1+\sigmavar^2)}}
  \frac{\sigmaperp}{\sigmaLN}
  ~\exp\left\{-\frac{[\ln(1+\nu\sigmavar)+\sigmaLN^2/2]^2}
        {2\sigmaLN^2}\right\}~~K(\gamma,\theta),
\\
\label{eq: N1_log}
 N_{1,\LN}^{(s)}(\nu,\theta) \,&=&\, \frac{1}{2\sqrt{\pi(1+\sigmavar^2)}}
  \frac{\sigmaperp}{\sigmaLN}
  ~\exp\left\{-\frac{[\ln(1+\nu\sigmavar)+\sigmaLN^2/2]^2}
        {2\sigmaLN^2}\right\}~~L(\gamma,\theta).
\end{eqnarray}
where the functions $J$, $K$ and $L$ are respectively given by 
(\ref{app_B: J}), (\ref{app_B: K}) and (\ref{app_B: L}) in Appendix 
\ref{appendix: result_JKL}, 
and we introduced the distortion parameter $\gamma$ given by 
\begin{equation}
 \label{eq: gamma}
      \gamma \equiv \sqrt{2}\,\frac{\sigmapara}{\sigmaperp}.    
\end{equation}

Notice that the parameter $\gamma$ equals unity when 
$\cpara=\cperp=1$ and the power spectrum is isotropic 
$P^{(s)}(k_{\scriptscriptstyle \|}, k_{\scriptscriptstyle \bot})=P^{(s)}(k)$
from equations (\ref{eq: sigmapara_P}) and (\ref{eq: sigmaperp_P}).  
When the geometric distortion is effective, the factors 
$\cpara$ and $\cperp$ are not equal to unity. Further, 
when the linear distortion or the finger-of-God effect
is influential, the power spectrum is no longer isotropic.  
Thus the deviation of $\gamma$ from unity quantifies 
the collected distortion effects, and represents anisotropies of the 
isodensity contours. 
As will be described in section \ref{subsec: results}, 
$\gamma$ is affected 
by the three distortion effects: the geometric distortion, 
the linear distortion, and the finger-of-God effect.
The linear distortion increases $\gamma$. On the contrary the 
finger-of-God effect decreases $\gamma$. 
As for the geometric distortion, this depends on the choice of 
cosmological redshift space $s(z)$ (see eqs.[\ref{eq: sigmapara_P}] 
and [\ref{eq: sigmaperp_P}]). 
When we adopt $s=\shb(z)$, $\gamma$ is decreased due to the 
geometric distortion. Conversely when adopting $s=\seds$, $\gamma$ 
is increased. These features of the geometric distortion are based
on the assumption that the real universe is the Lambda CDM model 
(see below). The strength of each distortion effect depends on 
smoothing scale and redshift, which yields the 
scale-/time-dependence of the value $\gamma$. 

\begin{figure}
\begin{center}
   \leavevmode\epsfxsize=12cm \epsfbox{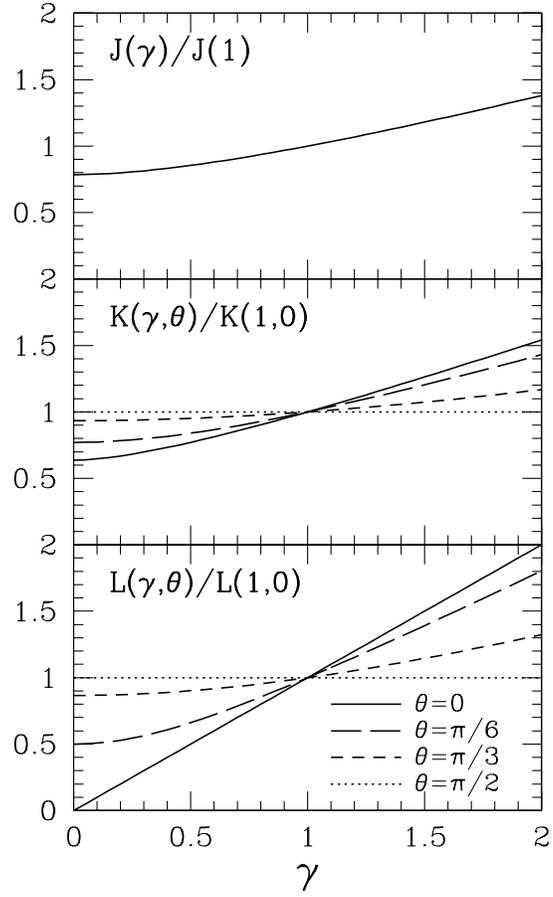}
\end{center}
\figcaption{ The normalized amplitude of 
$J$ ({\it Upper}), $K$ ({\it Middle}), and $L$ ({\it Lower}) 
as the function of the distortion parameter $\gamma$ 
(see also Appendix A). For $K$ and $L$ we fixed $\theta=0$ (solid), 
$\pi/6$ (long-dashed), $\pi/3$ (short-dashed), and $\pi/2$ (dotted). 
\label{fig: JKL}
}
\end{figure}

The above formulae state that the amplitude of the 
isodensity statistics depends on the geometric and the 
peculiar velocity effects, whose influences are 
characterized by the parameter $\gamma$. 
The sensitivity to those effects can be understood from the 
function $J$, $K$ and $L$. 
In figure \ref{fig: JKL}, we plot the normalized amplitude 
$J(\gamma)/J(1)$, $K(\gamma,\theta)/K(1,0)$ and $L(\gamma,\theta)/L(1,0)$ 
as the functions of distortion parameter $\gamma$. 
To show the angle dependence of the functions $K$ and $L$, 
we plot the cases $\theta=0$ ({solid}), $\pi/6$ ({long-dashed}), 
$\pi/3$ ({short-dashed}) and $\pi/2$ ({dotted}). 
{}Figure \ref{fig: JKL} suggests that the distortion effects 
may significantly affect the isodensity statistics, especially 
for the quantities described by the function $L(\gamma,\theta)$, 
i.e., $G_2$ and $N_1$. 
In next subsection, we will clarify how $\gamma$
is varied by the distortion effects at a quantitative level.

Before proceeding the analysis, it should be noted that 
isodensity statistics in terms of the volume fraction $f$ defined by 
\begin{equation}
  \label{eq: volume-frac}
  f\equiv\int_{\nu}^{\infty}\frac{dx}{\sqrt{2\pi}} \,\,e^{-x^2/2},  
\end{equation}
instead of the density threshold $\nu$, has been conventionally used in 
order to compare the Gaussian results with non-Gaussian cases 
(e.g, Gott et al. 1989; Vogely et al. 1994; Canavezes et al. 1998).
In fact, the isodensity statistics labeled by $f$ remains unchanged 
under the transformation (\ref{eq: del-phi}) in real space.  
In cosmological redshift space, however, the statistics cannot 
be entirely characterized by the volume fraction because of the 
anisotropy. Moreover, we intend to clarify influences of the 
distortion effects including the nonlinear correction.  
Therefore, we use the straightforward expressions labeled by 
the density threshold $\nu$. 
%
%
%
%
%
%
%
%
\subsection{Results} 
\label{subsec: results} 
%
%
%
%
%
%
Adopting the above nonlinear model, we now give a prediction for
isodensity statistics in specific cold dark matter (CDM) cosmological 
models. In this paper, we consider two typical CDM models, in which 
the cosmological parameters are chosen as 
$(\Omega_0,\lambda_0,\sigma_8,h)=(1.0,0.0,0.6,0.5)$ (SCDM model) and 
$(\Omega_0,\lambda_0,\sigma_8,h)=(0.3,0.7,1.0,0.7)$ (LCDM model),  
where $\sigma_8$, the top-hat mass fluctuation amplitude at $8h^{-1}$Mpc, 
is normalized by the cluster abundance (e.g, Kitayama \& Suto 1997). 
{}For the unknown biasing factor, we use the time-evolving biasing 
model by Fry (1996). This model simply accounts for the redshift 
evolution of number density fluctuations in the case that the 
formation and merging process can be negligible. 
Denoting the amplitude of the biasing factor at present by $b_{0}$, 
the linearized equation of continuity yields
   \begin{equation}
     \label{eq: FryBias}
     b(z)\,=\,1 + \frac{b_0-1}{D_+(z)}.
   \end{equation}
We fix the value $b_{0}=1.5$ for brevity.

\begin{figure}
\begin{center}

\vspace*{-2.0cm}

   \leavevmode\epsfxsize=10cm \epsfbox{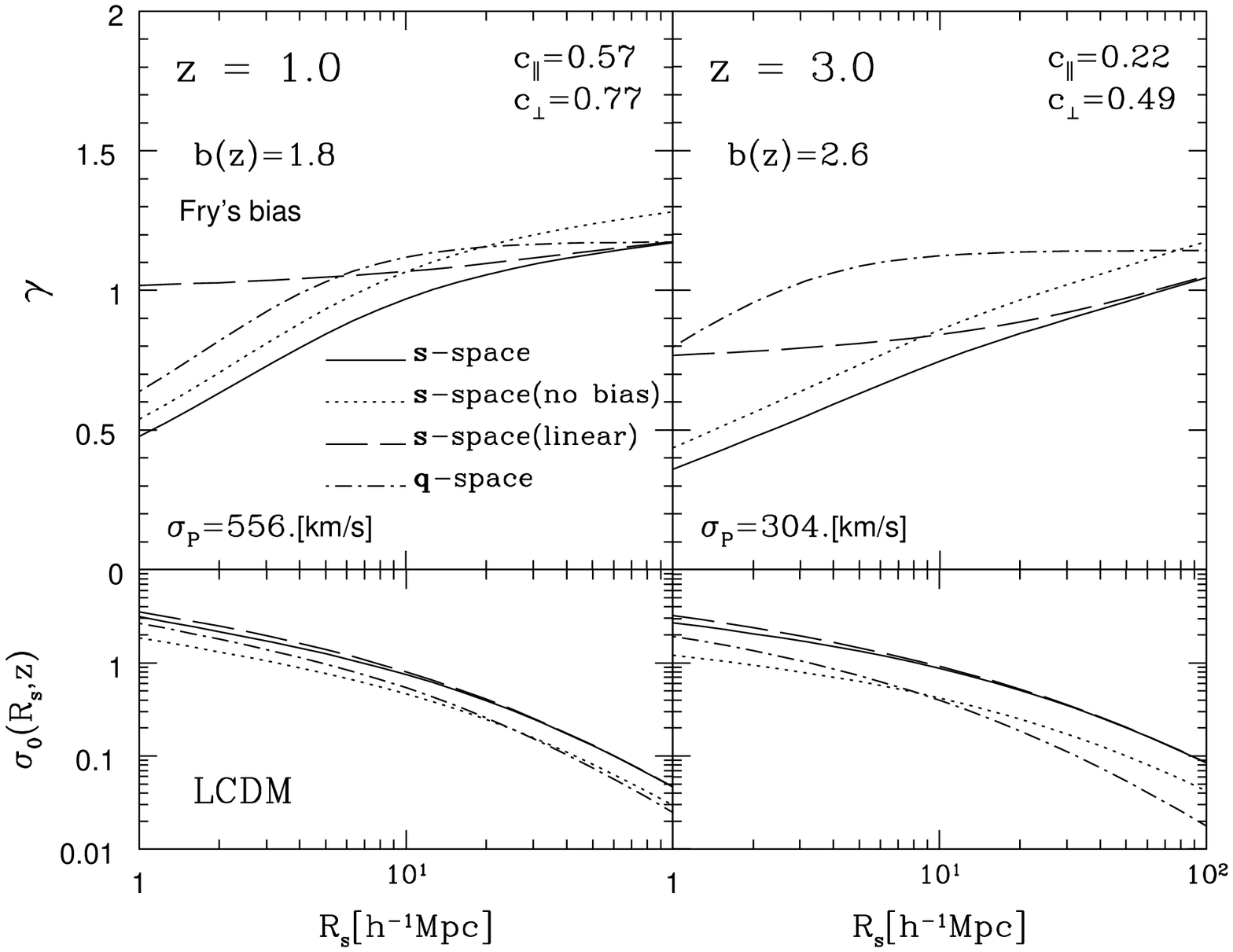}

\vspace*{-2.0cm}

   \leavevmode\epsfxsize=10cm \epsfbox{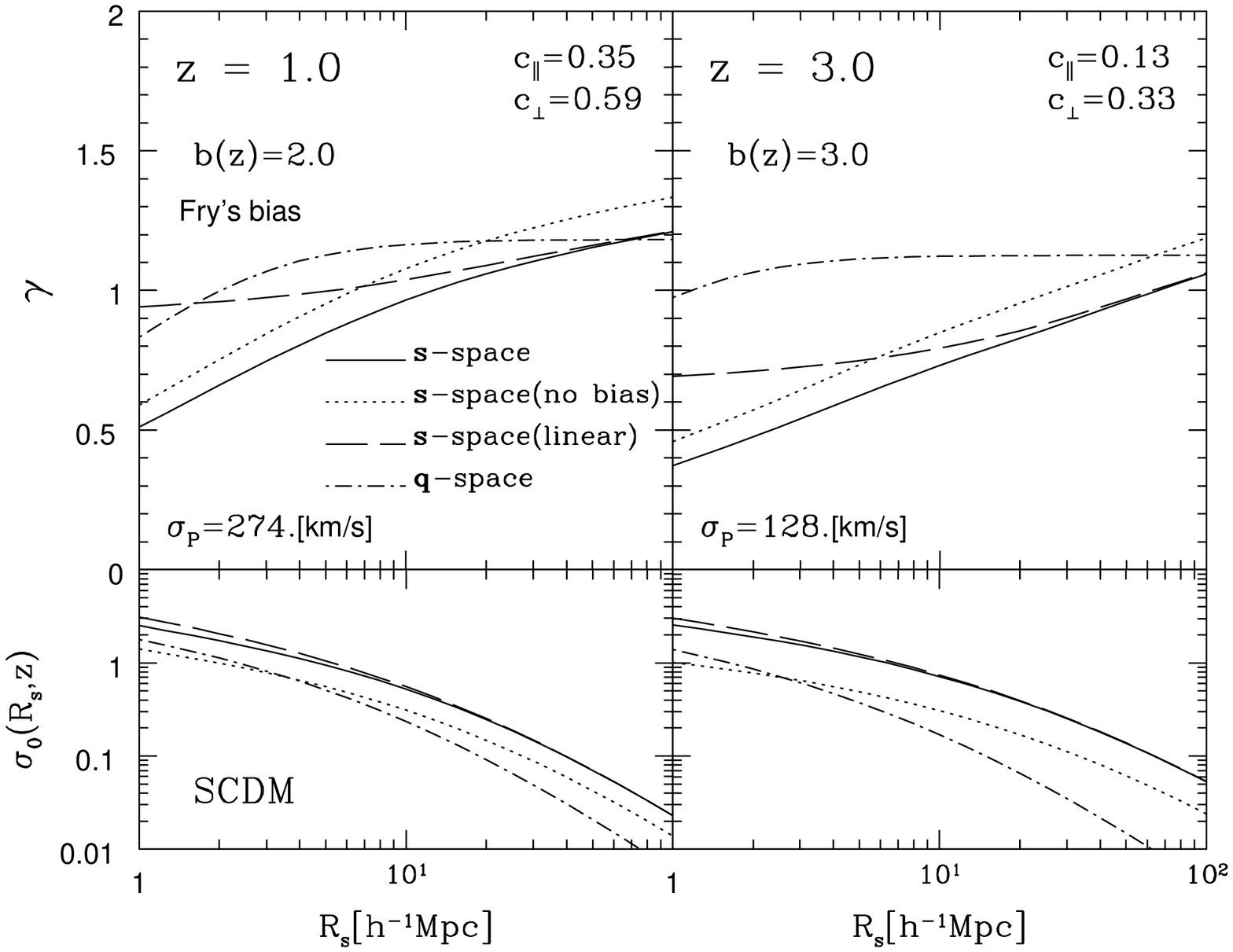}
\end{center}
\figcaption{ The scale-dependence of distortion parameter $\gamma$ 
and the field correlation $\sigmavar$ as the function of the
smoothing radius $R_{\rm s}$ in the case adopting the cosmological 
redshift space $s=\shb(z)$ (see eqs.[\ref{eq: shb}] and [\ref{eq: z_crd}]). 
{\it Upper-panels}({\it Lower-panels}) show the LCDM (SCDM) model.
The left (right) panel shows the case $z=1$ ($z=3$).
Here, we used the biasing parameter given by (\ref{eq: FryBias})
specifying the present value $b_0=1.5$. The solid and dot-dashed lines
indicate the predictions from the nonlinear power spectrum (\ref{eq:
redshift-P(k)}), where the latter case is obtained by setting 
$\cpara=\cperp=1$. The dashed line shows the linear prediction, 
in which only the linear distortion and the geometric effect are 
taken into account. For comparison, we plot the result for 
the no-biasing case, $b(z)=1$ (dotted line).   
\label{fig: gamma_Hubble}
}
\end{figure}

\begin{figure}
\begin{center}

\vspace*{-2.0cm}

   \leavevmode\epsfxsize=10cm \epsfbox{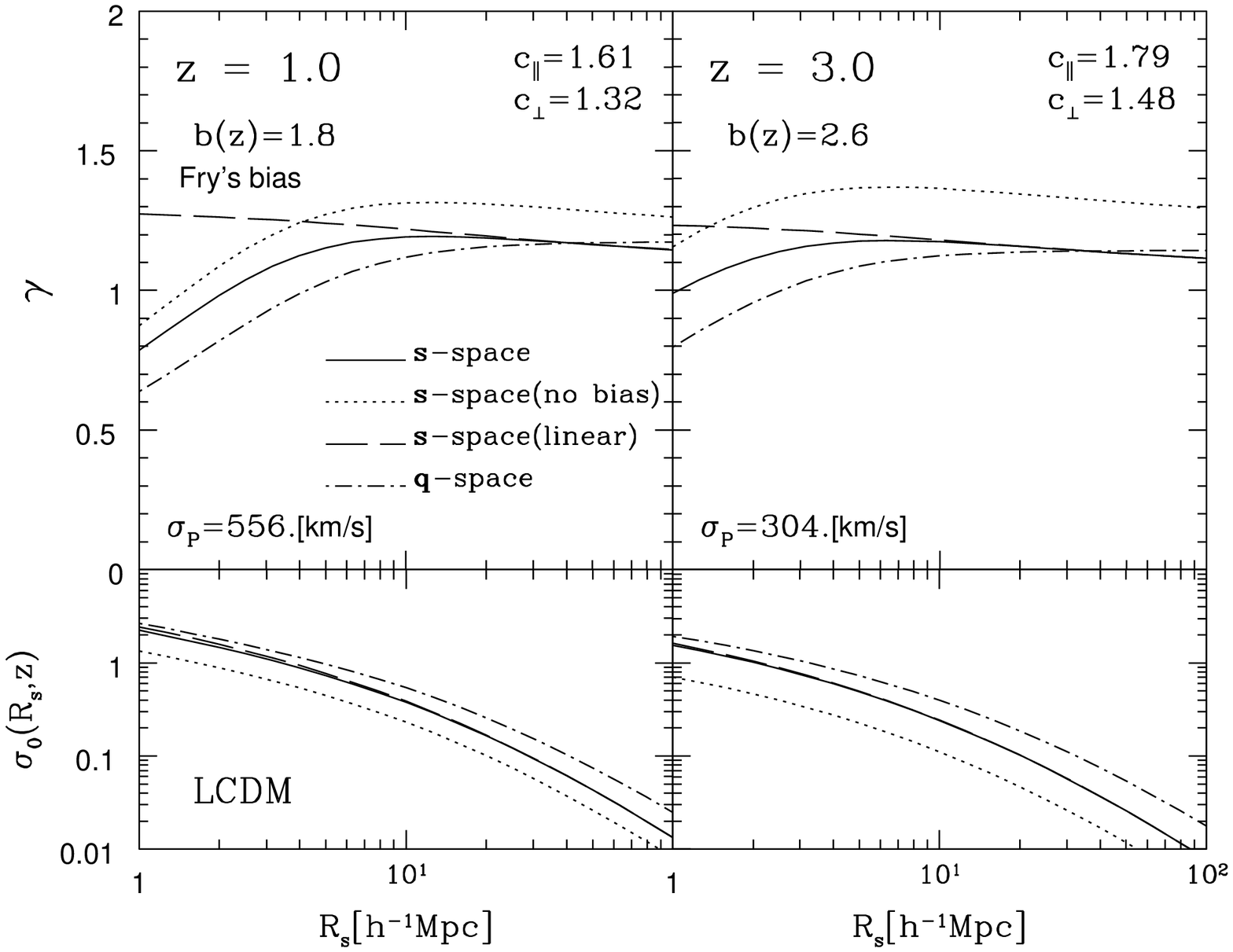}

\vspace*{-2.0cm}

   \leavevmode\epsfxsize=10cm \epsfbox{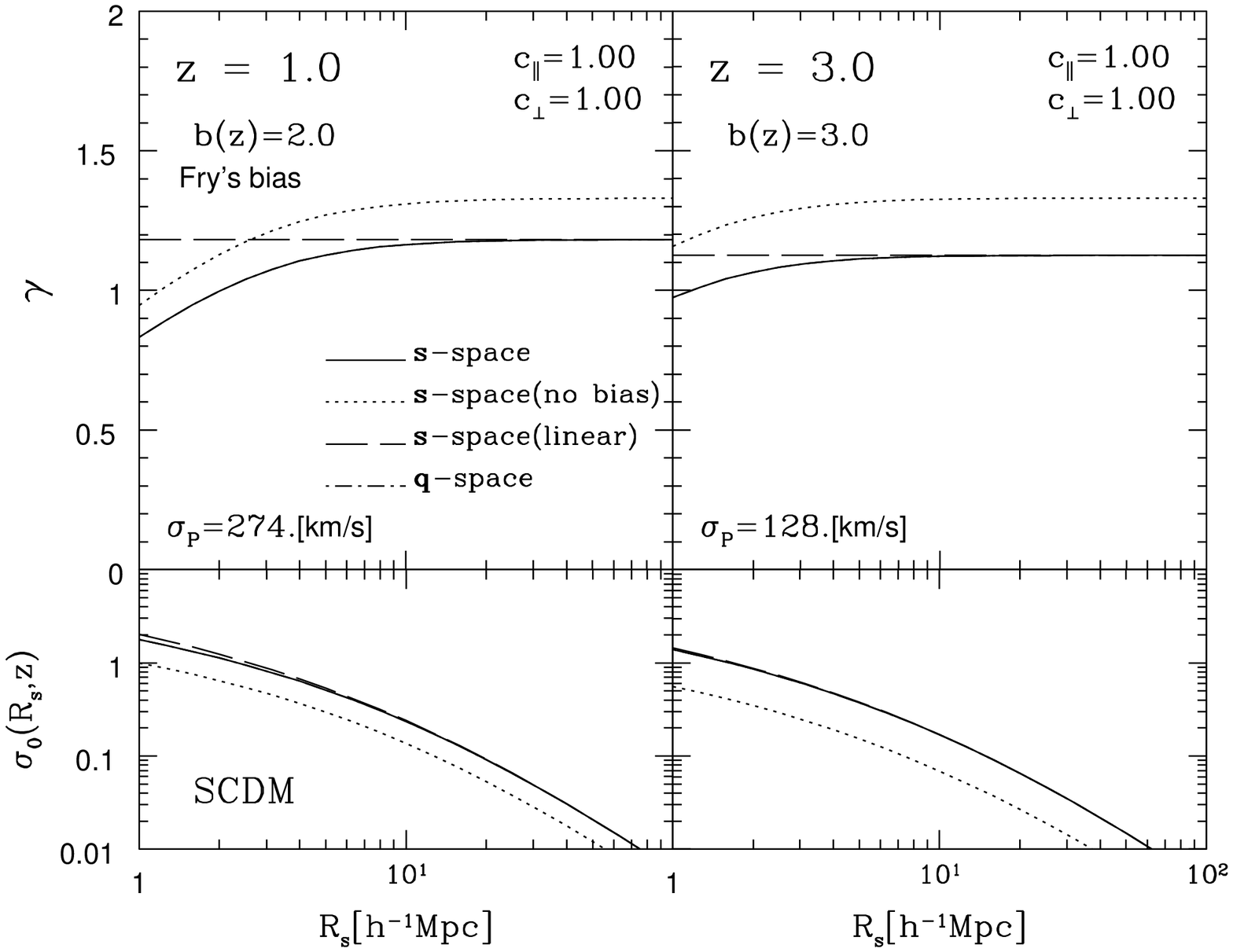}
\end{center}
\figcaption{ 
  Same as Figure \ref{fig: gamma_Hubble}, but we adopt the 
  cosmological redshift space 
  $s=\seds(z)$ (see eqs.[\ref{eq: seds}] and [\ref{eq: EdS_crd}]). 
  Note that the SCDM model (the lower panels) is not affected by 
  the geometric distortion because the cosmological redshift-space 
  is exactly same as the real space. Therefore the solid lines 
  exactly coincide with the dot-dashed lines in the SCDM model.
\label{fig: gamma_EdS}
}
\end{figure}

The distortion parameters $\gamma$ and the field correlation $\sigma_0$ 
are the most important parameters to understand behaviors of the shape
of the isodensity statistics. Therefore we first analyze $\gamma$ and 
$\sigmavar$. Figures \ref{fig: gamma_Hubble} and 
\ref{fig: gamma_EdS} show $\gamma$ and $\sigmavar$ as the functions 
of the smoothing radius $\rs$. Figure \ref{fig: gamma_Hubble} shows 
the case adopting the cosmological redshift space with $s=\shb(z)$ 
(see eqs.[\ref{eq: shb}] and [\ref{eq: z_crd}]). 
The results are depicted for the LCDM model 
({\it Upper-panel}) and the SCDM model ({\it Lower-panel}),
though the two models show almost similar behaviors.
In each model the results at $z=1$ ({\it Left-panel}) 
and $z=3$ ({\it Right-panel}) are shown.

On each panel the solid line plots the case all the effects are 
incorporated. Because the solid line corresponds to the case 
observed in the space $\s$, which is referred to '$\s$-space' on 
the panel. The solid line shows that the distortion parameter 
$\gamma$ (the clustering amplitude $\sigmavar$) gradually increases 
(decreases) as the smoothing scale $\rs$ increases.  
To understand the importance of the nonlinear and the geometric 
distortion effects, we also plot the linear results neglecting the 
nonlinear correction ({\it dashed line}) and the case in absence of 
the geometric distortion ({\it dot-dashed line}), which are referred to 
'$\s$-space(linear)' and '$\q$-space' on the panel, respectively.
The latter case corresponds to the result obtained in the $\q$-space 
(see eq.[\ref{eq: s_vs_x}]), which is the same result obtained in 
the $\s$-space with setting $\cpara=\cperp=1$.

By comparing the behaviors of the solid and the dot-dashed 
lines for $\gamma$ on large smoothing scales, where the linear 
distortion and the geometric distortion are influential, 
it is apparent that the linear distortion increases $\gamma$ 
and that the geometric distortion decreases $\gamma$. 
On the other hand the nonlinear correction becomes important 
on small scales (compare {\it solid} and {\it dashed} lines).  
It appears that the nonlinear effects cause more significant 
influence on the distortion parameter $\gamma$ rather than the 
clustering amplitude $\sigmavar$. 
This is because the finger-of-God effect compensates 
the gravitational nonlinear growth and the resultant $\sigmavar$ 
becomes almost same as the linear result. As for $\gamma$, since 
the finger-of-God effect dominates the linear velocity distortion
on small scales, $\gamma$ is significantly decreased.
It is notable that the correction from nonlinear effects still 
remains non-negligible even at the larger smoothing length 
$\rs\sim20h^{-1}$Mpc in the cosmological redshift space $\shb$.  
This behavior is caused from the scaling effect of the geometric
distortion through the factors $\cpara$ and $\cperp$. 
Note also that the difference between the solid line and the
dot-dashed line becomes significant at higher redshift $z=3$,
which also indicates the importance of the geometric distortion 
effect at the higher redshift.

Figure \ref{fig: gamma_EdS} shows another example, where the cosmological 
redshift space is chosen as $s=\seds(z)$ (see eqs.[\ref{eq: seds}] 
and [\ref{eq: EdS_crd}]) . 
In contrast with the case  $s=\shb(z)$ 
in Figure \ref{fig: gamma_Hubble}a, the geometric distortion
increases $\gamma$ and suppresses the amplitude $\sigmavar$ 
in the case of the LCDM model ({\it Upper-panel}). 
{}For the SCDM case, the results in $\s$-space exactly coincides with 
those in $\q$-space because the cosmological redshift space 
coincide with the comoving real space and the geometric distortion is absent. 
In both cases, correction from the nonlinear effects becomes 
negligible on the scales $\rs \simgt 8 h^{-1}$Mpc at high-redshift.  
This is also true irrespective of whether the cosmological model 
of the real space is spatially flat or open.

For comparison, in Figure \ref{fig: gamma_Hubble} and \ref{fig: gamma_EdS}, 
the case of no bias is also shown by setting $b(z)=1$ 
({\it dotted line}). 
This alternation 
does not change the qualitative behavior, however, we can see that 
$\gamma$ and $\sigmavar$ are affected by the amplitude
of the bias as follows. Increasing the amplitude of the biasing 
factor, $\gamma$ is decreased, conversely $\sigmavar$ is increased. 
This effect could be more significant at higher redshift 
because of time-evolution of the bias.

\begin{figure}
\begin{center}
   \leavevmode\epsfxsize=7.2cm \epsfbox{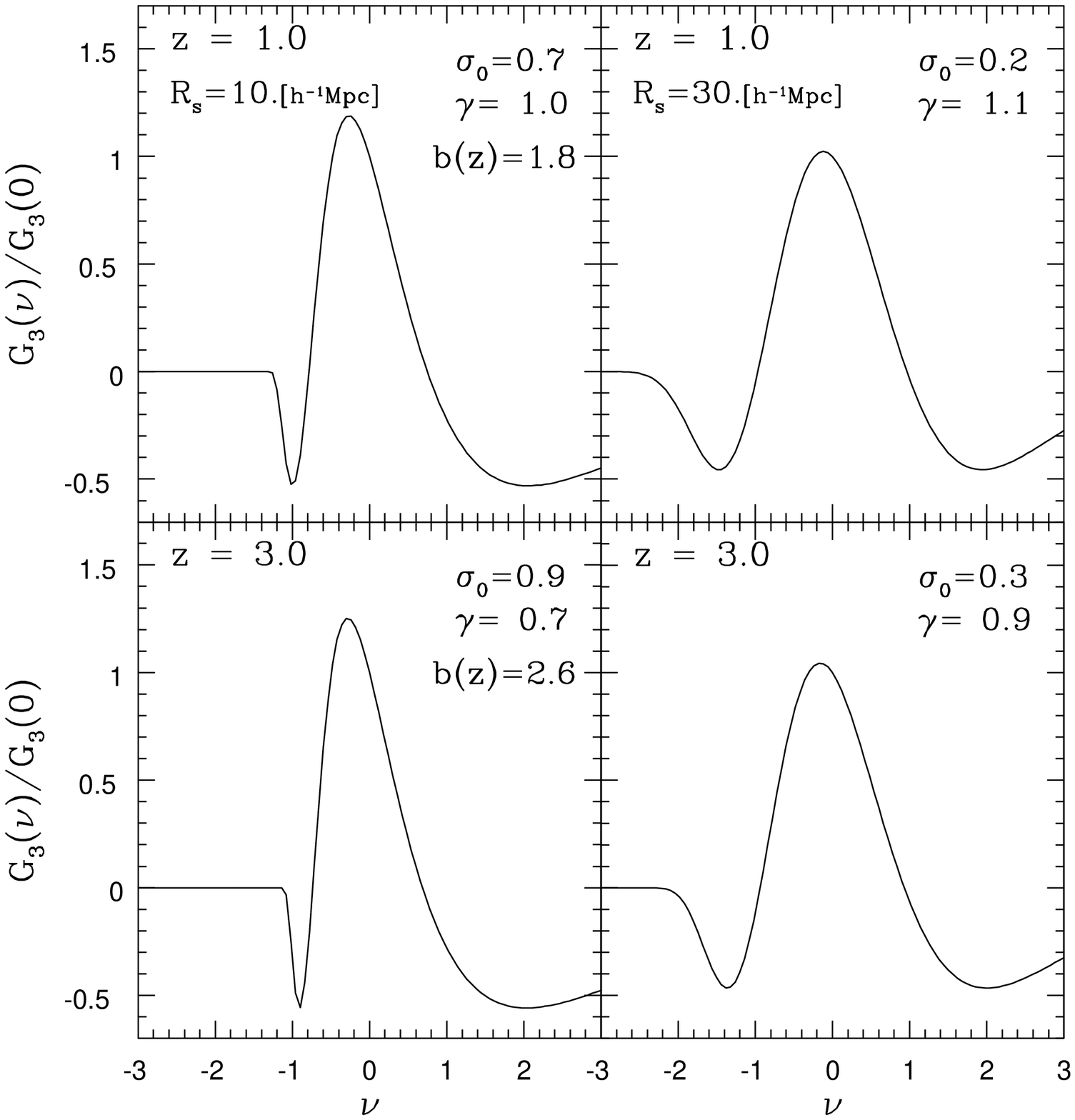}
   \leavevmode\epsfxsize=7.2cm \epsfbox{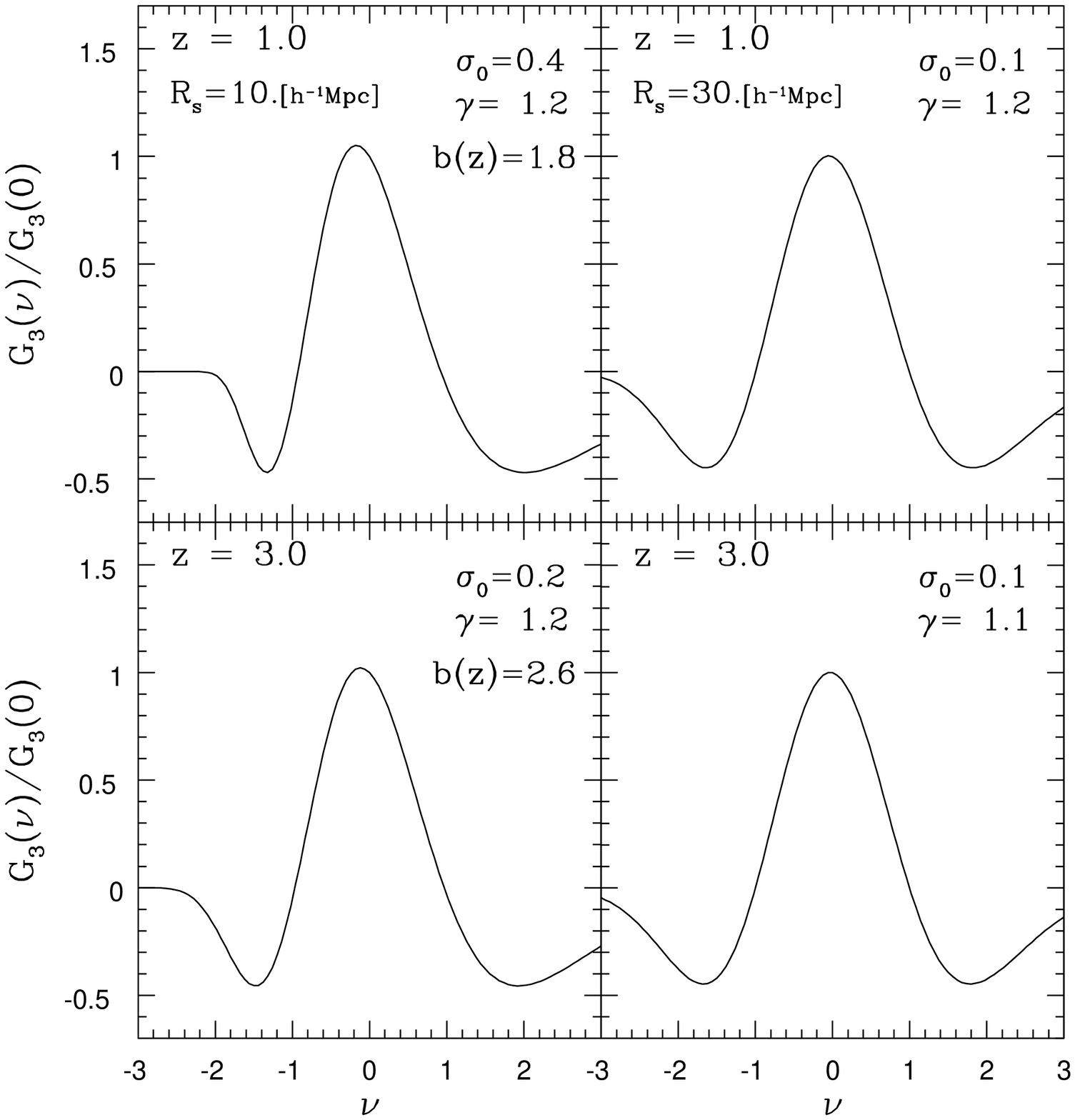}\\
   \leavevmode\epsfxsize=7.2cm \epsfbox{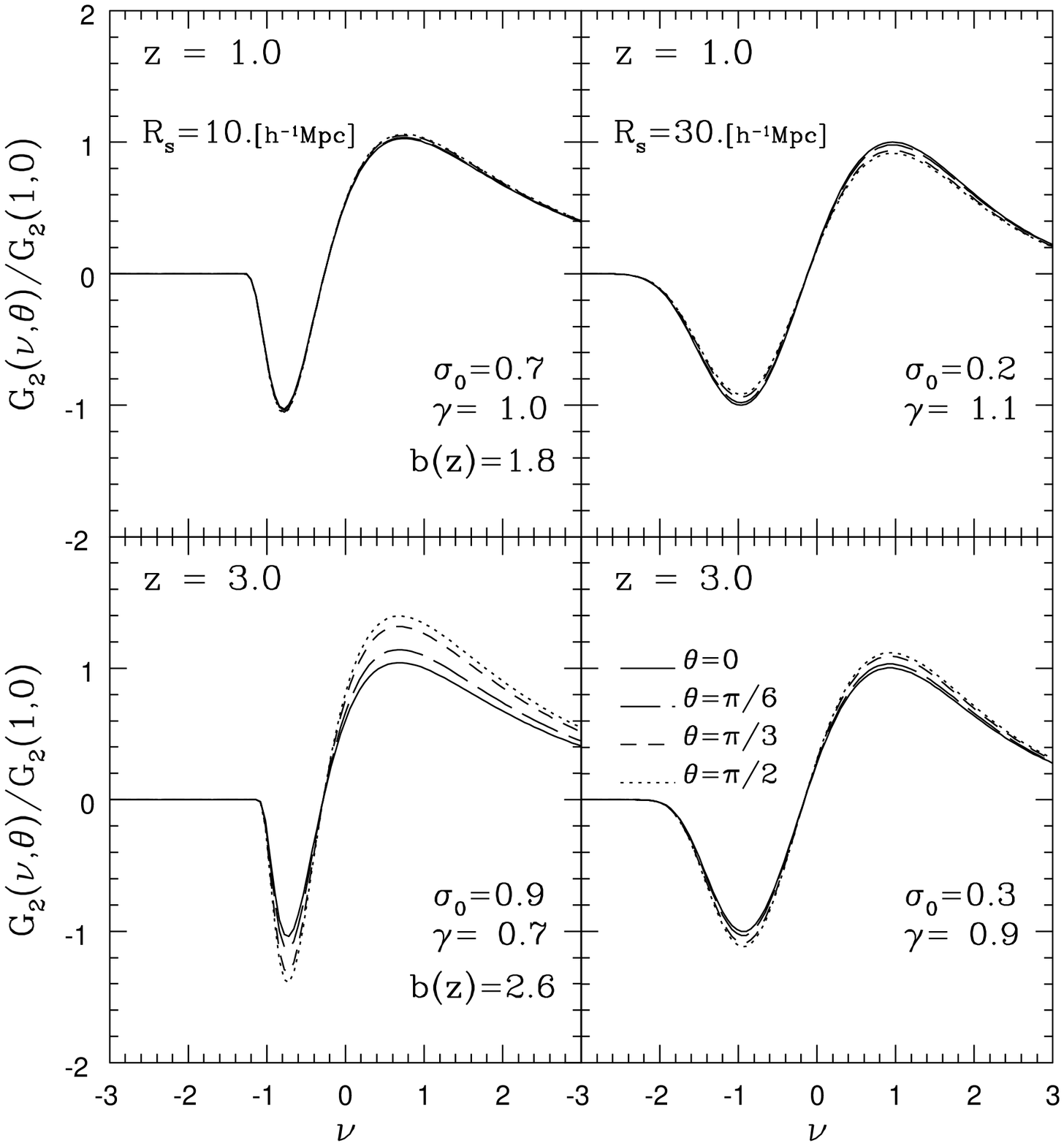}
   \leavevmode\epsfxsize=7.2cm \epsfbox{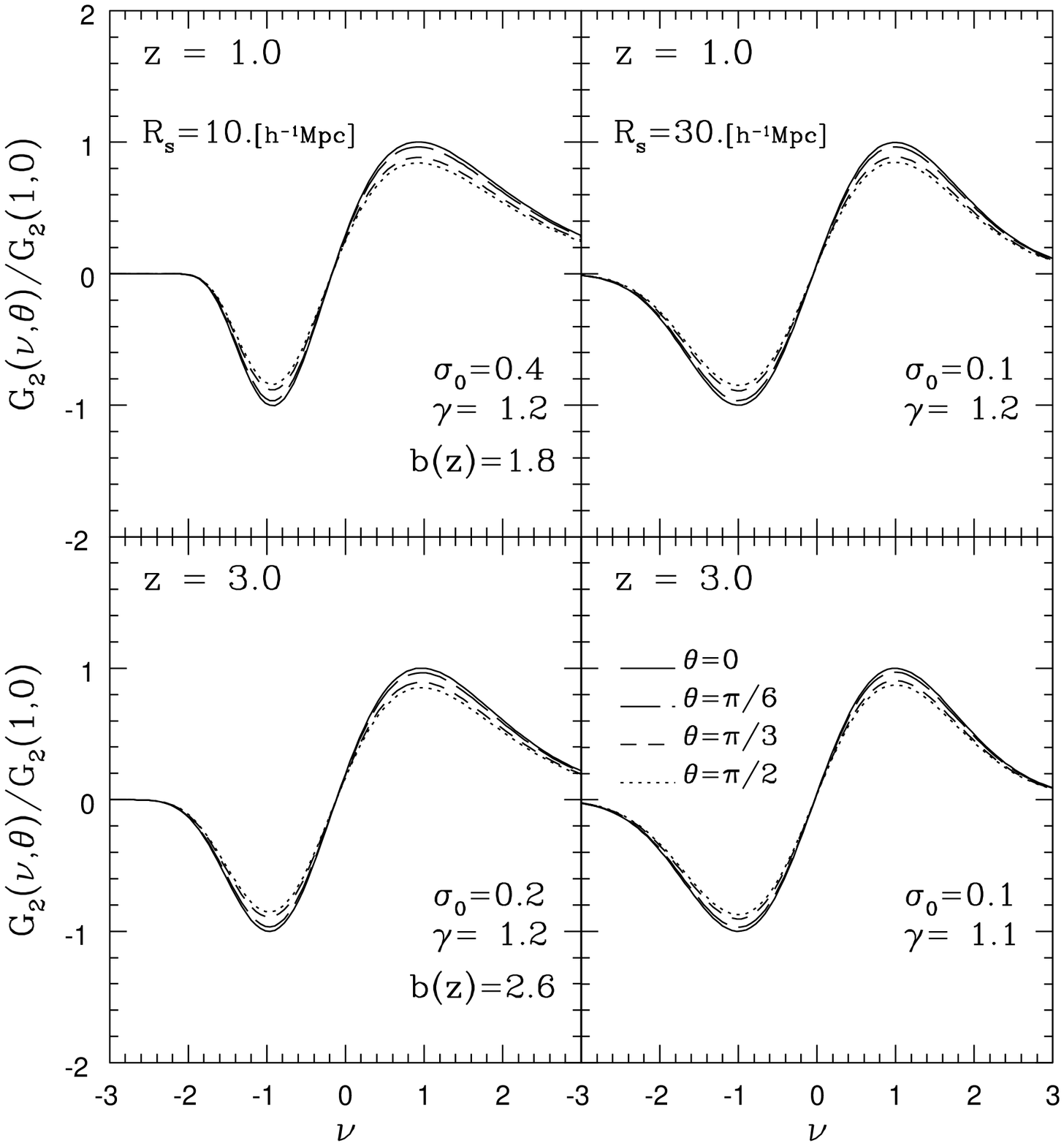}\\
\end{center}
\figcaption{Shapes of the isodensity statistics as the function 
of the density height $\nu$ in the LCDM model. We assume the 
cosmological redshift space $s=\shb(z)$ in left-panel, and 
$s=\seds(z)$ in right-panel. In each panel, we show the result 
with the redshift and the smoothing scale fixed as
$z=1$ and $\rs=10h^{-1}$Mpc (upper-left), 
$z=1$ and $\rs=30h^{-1}$Mpc (upper-right), 
$z=3$ and $\rs=10h^{-1}$Mpc (lower-left), and 
$z=3$ and $\rs=30h^{-1}$Mpc (lower-right).    
As for the biasing parameter, we adopt the Fry's bias model 
with the present value $b_0=1.5$ : the three-dimensional genus 
$G_3$ ({\it Upper}); the two-dimensional genus $G_2$ ({\it Lower}). 
{}For $G_{2}$, we plot the cases $\theta=0$ (solid), 
$\pi/6$ (long-dashed), $\pi/3$ (short-dashed), and $\pi/2$ (dotted).
\label{fig: GNcurve1}
}
\end{figure}

\begin{figure}
\begin{center}
   \leavevmode\epsfxsize=7.2cm \epsfbox{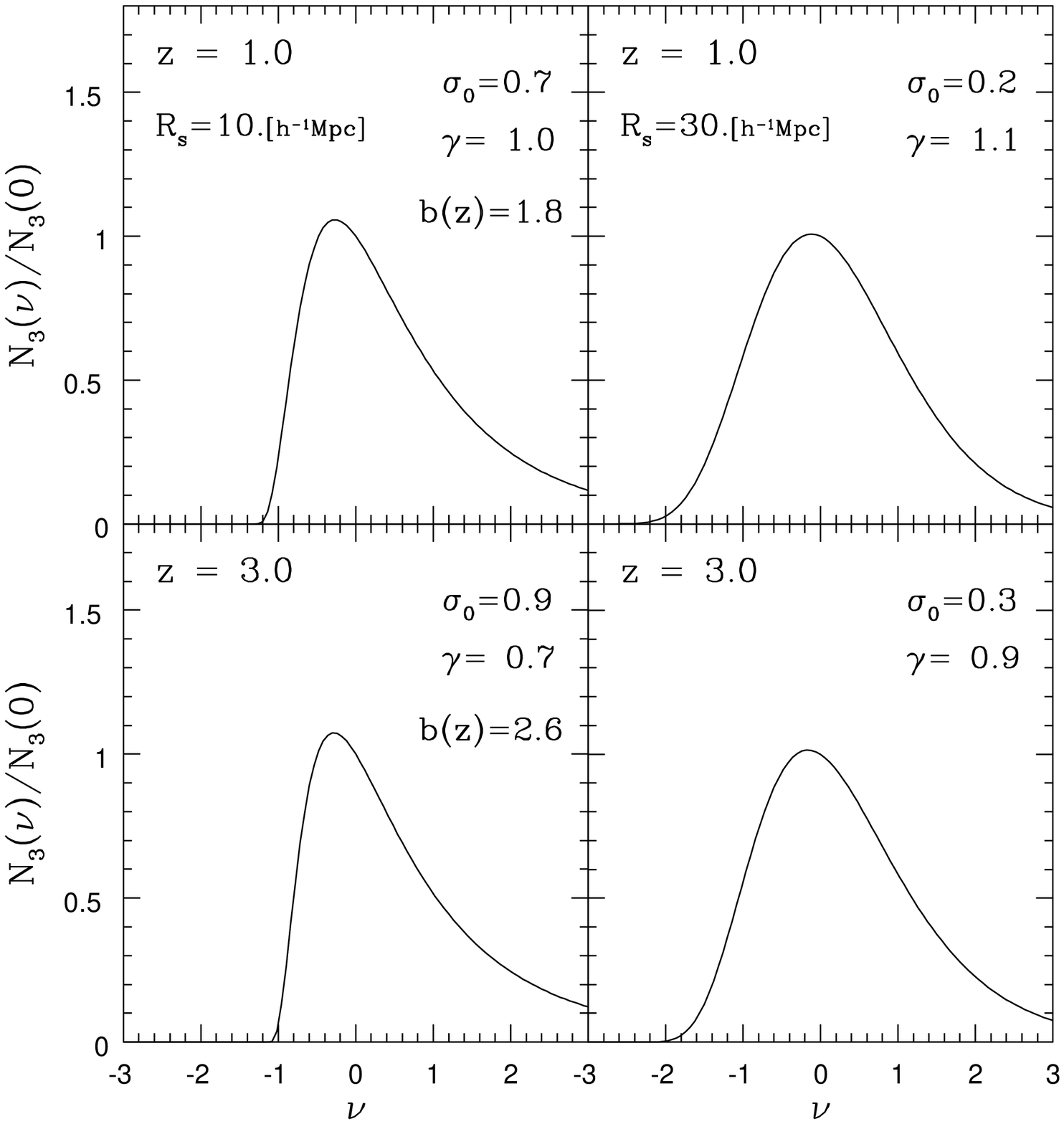}
   \leavevmode\epsfxsize=7.2cm \epsfbox{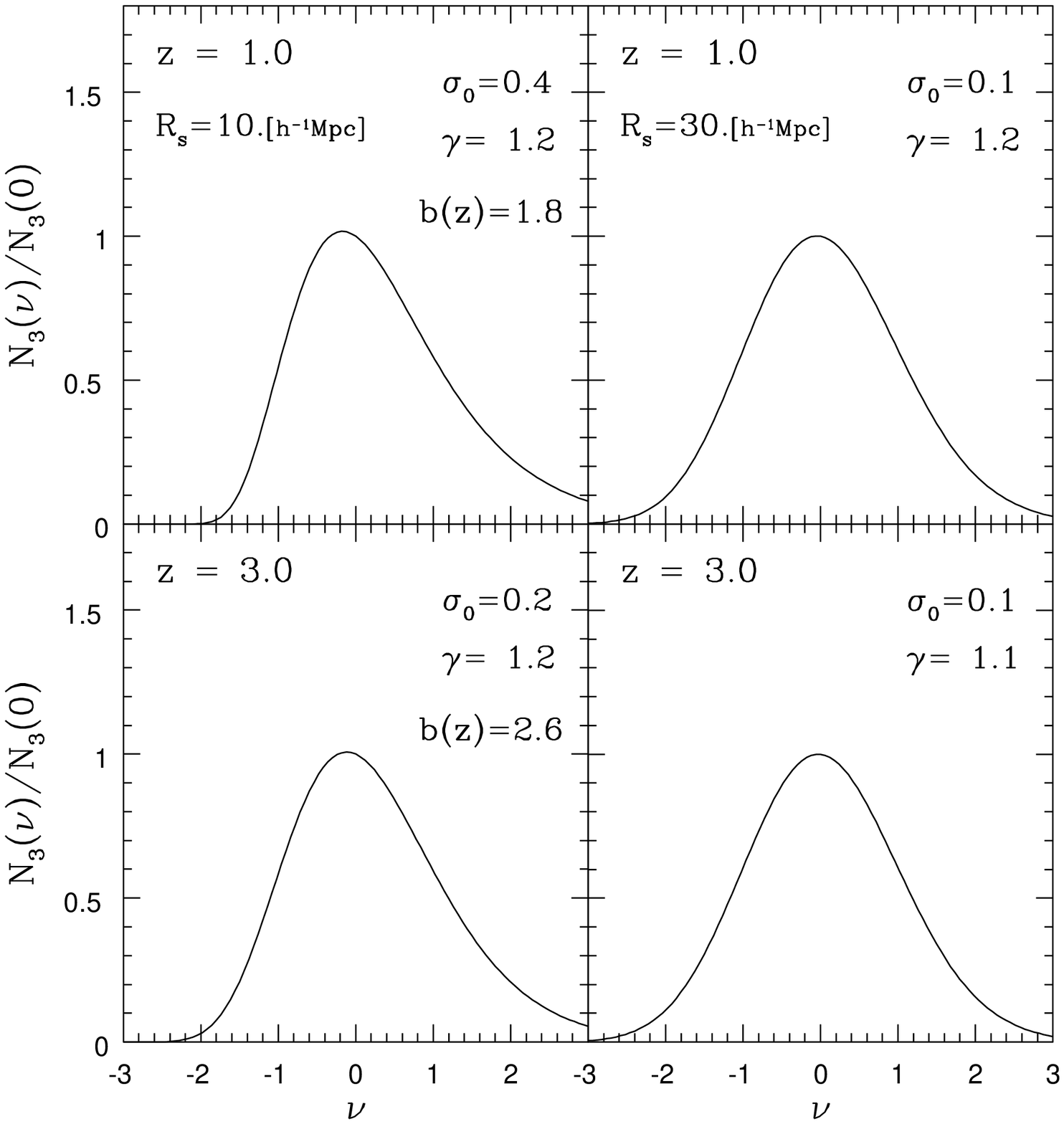}\\ 
   \leavevmode\epsfxsize=7.2cm \epsfbox{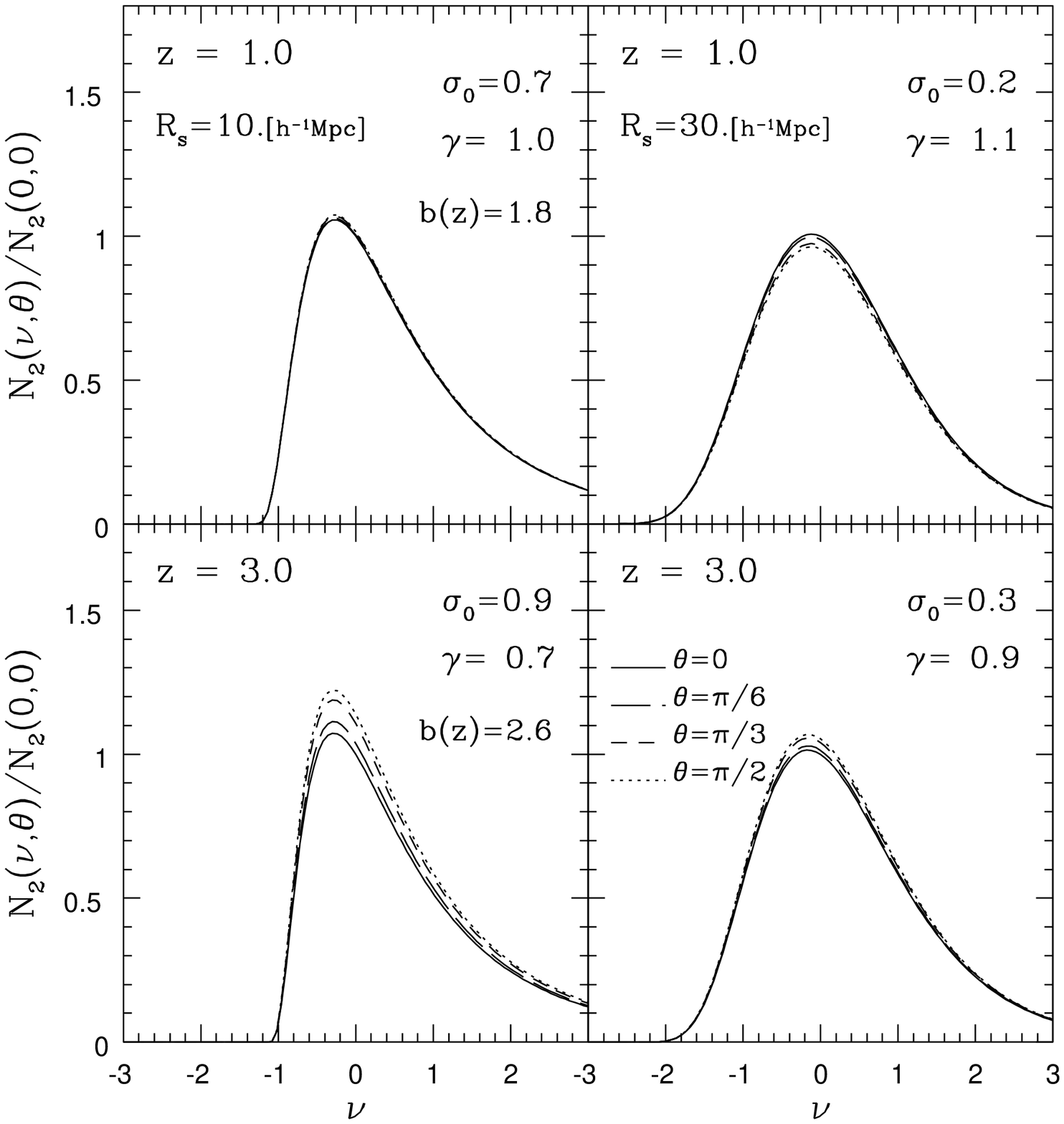}
   \leavevmode\epsfxsize=7.2cm \epsfbox{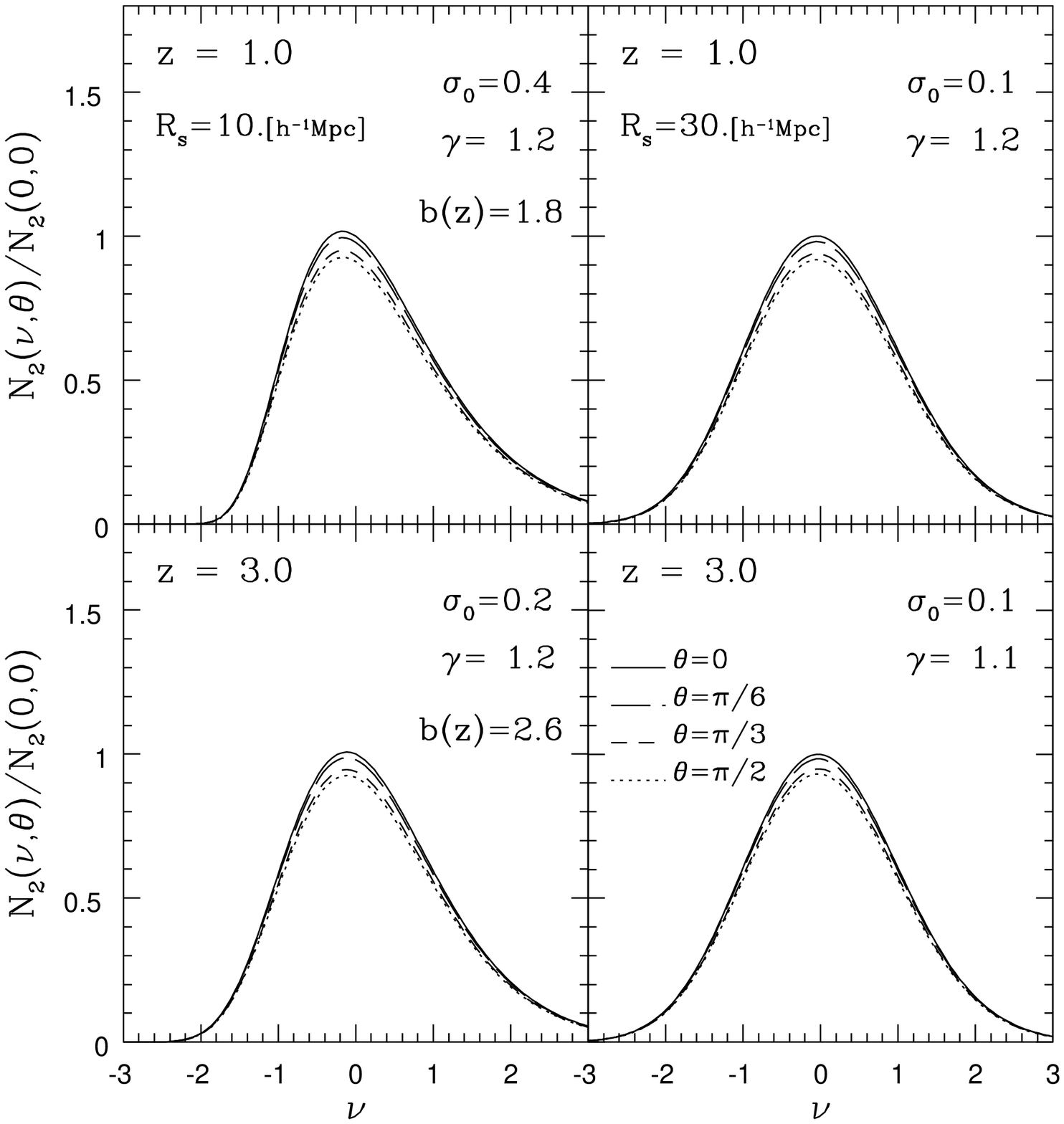}\\
   \leavevmode\epsfxsize=7.2cm \epsfbox{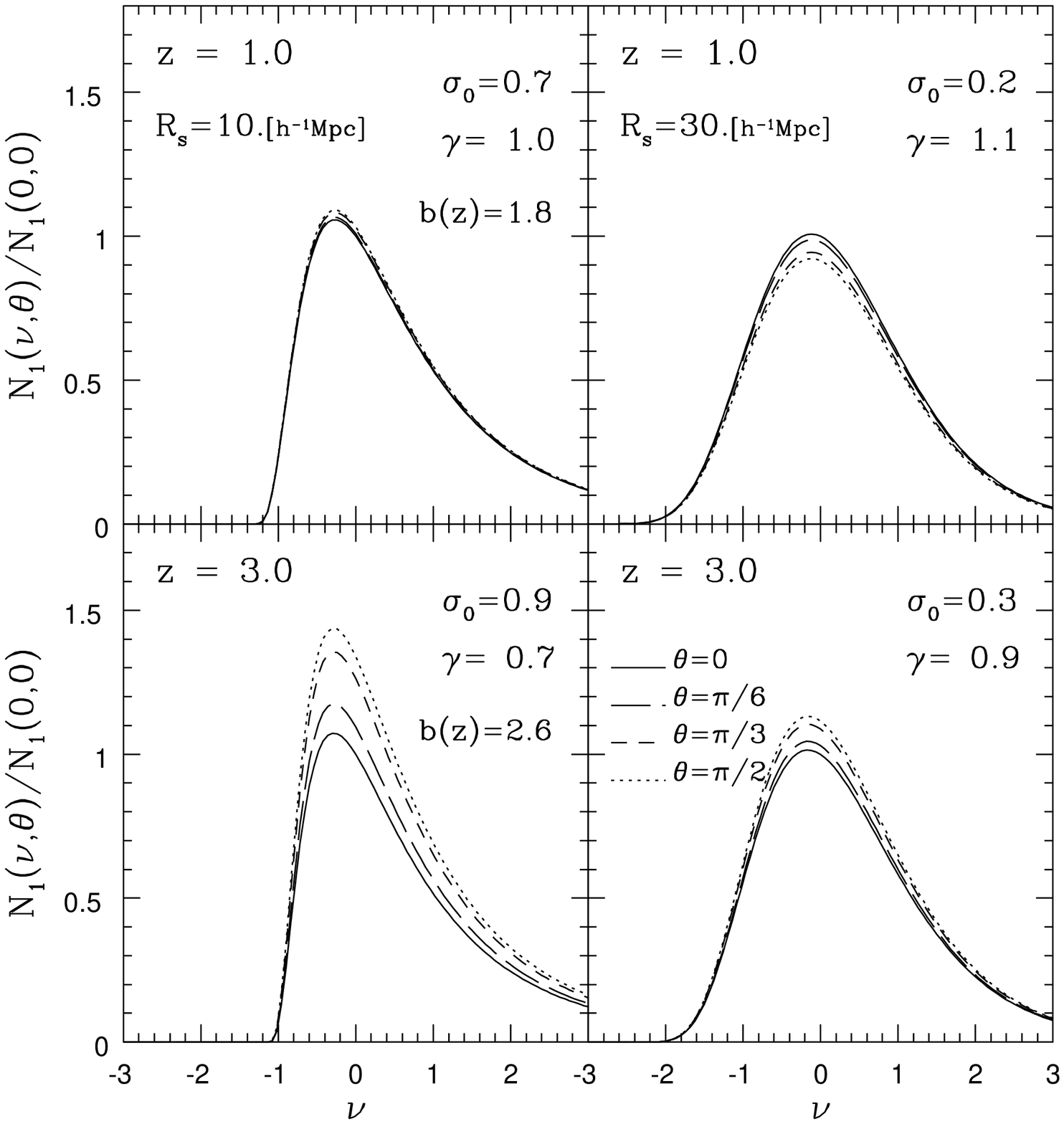}
   \leavevmode\epsfxsize=7.2cm \epsfbox{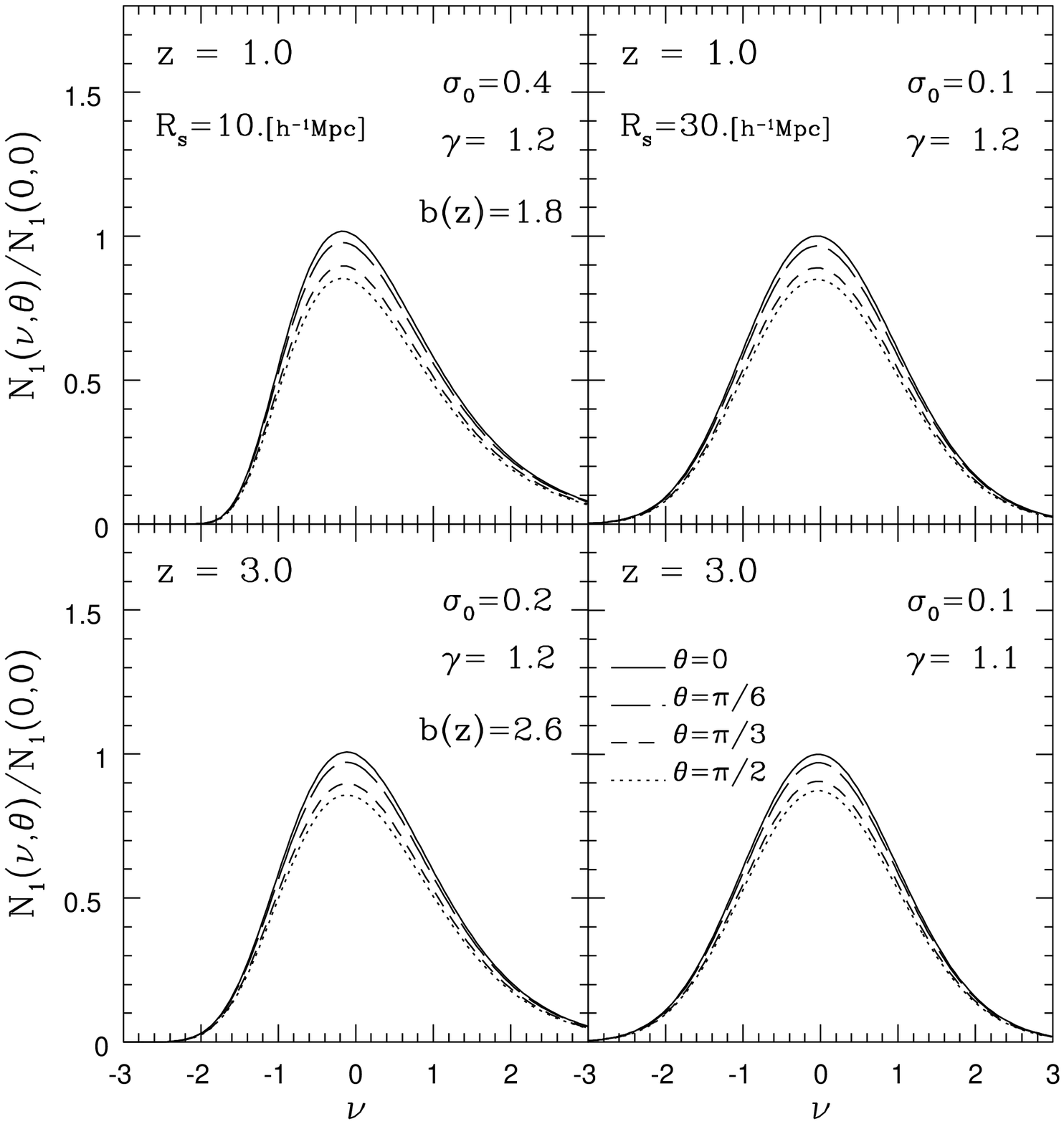}
\end{center}
\figcaption{
  Same as Figure \ref{fig: GNcurve1}, but we plot 
  the mean area of isodensity contour $N_{3}$ ({\it Upper}), 
  the mean length of isodensity contour intersecting 
  with a plane $N_{2}$ ({\it Middle}), and 
  the mean number of crossing points of isodensity contour 
  on a line $N_{1}$ ({\it Lower}). For $N_{2}$ and $N_{1}$,
  we plot the cases $\theta=0$, $\pi/6$, $\pi/3$, and $\pi/2$.
\label{fig: GNcurve2}
}
\end{figure}

In principle, behaviors of the isodensity statistics can be entirely 
understood from Figure \ref{fig: JKL}, \ref{fig: gamma_Hubble}, 
and \ref{fig: gamma_EdS}. 
As a demonstration, however, in Figure \ref{fig: GNcurve1} we plot 
shape-dependence of the genus $G_3$~({\it upper panels}) and 
$G_2$~({\it lower panels}) as functions of the density height 
$\nu$ for the LCDM model. 
{}Figure \ref{fig: GNcurve2} shows the area statistics 
$N_3$~({\it upper panels}), $N_2$~({\it middle panels}), and 
$N_1$~({\it lower panels}).
In both Figures \ref{fig: GNcurve1} and \ref{fig: GNcurve2}, 
the cosmological redshift space $s=\shb(z)$ ($s=\seds(z)$) is 
chosen for the panels in the left (right) hand side of the figures.
As is expected, the shape of the isodensity statistics is very 
sensitive to the choice of the cosmological redshift space. 
In the case $s=\shb(z)$, the significance of the nonlinear correction 
leads to asymmetric shapes of the isodensity statistics even at the 
large smoothing length $\rs\simeq 10h^{-1}$Mpc and at high-redshift.   
The relative amplitude of $G_2$ and $N_1$ is sensitive to the angle 
$\theta$, which might be a useful tool to determine the 
cosmological model of our universe as a simple geometric test 
(Alcock \& Paczy\'nski 1978; see Sec.\ref{sec: conclusion}). 
On the other hand, in the case $s=\seds(z)$, 
all the statistics have nearly symmetric shape. 
This indicates that the Gaussian 
linear results could be a good approximation. Interestingly, at 
high-redshift $z=3$, the relative amplitude of the isodensity 
statistics has the opposite $\theta$-dependence in the case 
$s=\seds(z)$, compared to the one in the case $s=\shb(z)$.   
As the angle $\theta$ increases, the relative amplitude becomes 
smaller in the case $s=\seds(z)$ and conversely larger in the 
case $s=\shb(z)$ (c.f. {\it left}- and {\it right-panels}). 
This behavior comes from the difference of the value $\gamma$.
{}For the smoothing length adopted here, $\gamma<1$ in the case
$s=\shb(z)$, while $\gamma>1$ in the case $s=\seds(z)$ 
(Figures \ref{fig: gamma_Hubble} and \ref{fig: gamma_EdS}).
This fact causes the opposite $\theta$-dependence, as 
is easily deduced from Figure \ref{fig: JKL}.

\section{Discussion}
\label{sec: discussion}
%
%
%
%
%
%
The analysis in previous section shows that the anisotropy caused by 
the geometric and the velocity distortions systematically affect 
the amplitude of the one- and two-dimensional quantities with the 
fractional power by $10\sim 20\%$.  
For the three-dimensional quantities $G_3$ and $N_3$, the 
scaling effect by the geometric distortion enhances or reduces the 
nonlinearity. These results are qualitatively correct as long as a 
large number of samples are distributed within a narrow range of 
the redshift. In this section, for practical purpose, we focus on 
the following issues: detectability of these effects 
(subsection \ref{subsec: sensitivity}); influence of finite 
redshift range of samples, i.e, the light-cone effect 
(subsection \ref{subsec: light-cone}). 
%
%
%
%
%
%
%
%
%
%
\subsection{Sensitivity}
\label{subsec: sensitivity}
%
%
%
For quantitative study of high-redshift clustering, 
we must be careful for the Poisson noise and the sampling noise as the 
main statistical errors. This is not an exception for the case of 
isodensity statistics. While the sampling noise arises from finiteness 
of the sampling volume, the Poisson noise masks the 
isodensity statistics of underlying density fluctuations 
when the samples of cosmological objects are sparse and rare.   
In this subsection, to compare the errors with the systematic effects 
of geometric and redshift-space distortions, we roughly estimate both 
errors in an analytical manner, and investigate sensitivity 
of the isodensity statistics assuming typical high-redshift samples.

\def\G{{\cal G}}
Let us first estimate the Poisson noise. For simplicity, we consider 
the isodensity 
statistics for the Gaussian random fields. 
The analytic formulae in Appendix \ref{appendix: gaussian results} 
reveal that the amplitude of the isodensity statistics can be
written in a general form :
\begin{equation}
 \G_i^{(s)}(\nu)\,\,\,\propto\,\,\,
  \left[\frac{\langle\,\, \ks^2 \,\,\rangle}{3}\right]^{i/2}, 
\label{eq: amplitude}
\end{equation}
where $i=3$ for $G_3$, $i=2$ for $G_2$, and $i=1$ for the area
statistics, and we defined
\begin{equation}
  \label{eq: k-mode_correlation}
\langle\,\, \ks^2 \,\,\rangle \,\,\,\equiv \,\,\,
\frac{\displaystyle \int d^3\K_s\,\,
\hat{W}^2(\ks\rs) P^{\rm \scriptscriptstyle (crd)}(\ks)\,\,\ks^2}
{\displaystyle \int d^3\K_s\,\,
\hat{W}^2(\ks\rs) P^{\rm \scriptscriptstyle (crd)}(\ks)}, 
\end{equation}
with the Gaussian window function $\hat{W}(y)=e^{-y^2/2}$. 
In the expression (\ref{eq: amplitude}), we have simply ignored the 
anisotropy due to the redshift-space distortions.  

The influence of the Poisson noise can be roughly evaluated as follows.
The effect of the Poisson sampling is incorporated by adding the
shot-noise power spectrum (e.g., Feldman, Kaiser \& Peacock 1994) 
\begin{eqnarray}
  &&P^{\rm \scriptscriptstyle (crd)}_{\rm \scriptscriptstyle shot}
  =\frac{1}{n_s}, 
\end{eqnarray}
where $n_s$ is the mean number density of the observed high-redshift 
objects. Then we insert the following power spectrum 
$P^{\rm \scriptscriptstyle (crd)}(\ks)$ into equation 
(\ref{eq: k-mode_correlation}),
\begin{eqnarray}
 \label{eq: power_poisson}
  &&P^{\rm \scriptscriptstyle (crd)}(\ks) = 
  P^{\rm \scriptscriptstyle (crd)}_{\rm \scriptscriptstyle true}(\ks) + 
  P^{\rm \scriptscriptstyle (crd)}_{\rm \scriptscriptstyle shot}~,
\end{eqnarray}
where $P^{\rm \scriptscriptstyle (crd)}_{\rm \scriptscriptstyle true}(\ks)$ 
means the spectrum free from the shot noise. By substituting equation 
(\ref{eq: power_poisson}) into (\ref{eq: amplitude}), we can evaluate 
the ratio of $\G_i^{(s)}$ incorporating the Poisson 
term to the one neglecting it (Canavezes et al. 1998). 
Here we consider the case that the spectrum is given in the power low 
form $P^{\rm \scriptscriptstyle (crd)}_{\rm \scriptscriptstyle true}
\propto k^{-1}$ in the cosmological redshift space for simplicity. 
This yields the correlation function of the form 
\begin{eqnarray}
      \xi^{(s)}(s)=\int \frac{d^3\K_s}{(2\pi)^3}\,\,
      P^{\rm \scriptscriptstyle (crd)}_{\rm \scriptscriptstyle true}(\ks)\,\,
      \frac{\sin(\ks s)}{\ks s}\,\, =\,\, \left(\frac{s}{s_0}\right)^{-2}, 
\nonumber
\end{eqnarray}
where $s_0$ is a correlation length. Then 
the ratio of $\G_i^{(s)}$ reduces to 
\begin{eqnarray}
 \frac{\G_i^{(s)}+\Delta \G_i^{(s)}}{\G_i^{(s)}}\,\,&=& \,\,
  \left(\frac{\displaystyle ~~1~+~\frac{3}{8\pi^{3/2}}\,\,
   \frac{1}{{n_s}\,s_0^2\, \rs}~~}
  {\displaystyle ~~1~+~\frac{1}{4\pi^{3/2}}\,\,
  \frac{1}{{n_s}\,s_0^2\, \rs}~~}\right)^{i/2}
\,\, 
  \simeq\,\,
  \left(1+\frac{1}{8\pi^{3/2}}\,\,\frac{1}{{n_s}\,s_0^2\,\rs}\right)^{i/2}, 
\label{eq: ratio_of_amp}
\end{eqnarray}
where the last line is valid as long as the contribution from 
the Poisson noise term is small compared with the dominant 
contribution from the Gaussian random field. 
Thus the quantity defined by 
\begin{equation}
\label{eq: S/N}
 \left(\frac{S}{N}\right)_{\rm shot}\,\,\,\equiv \,\,8\pi^{3/2}\,
 {n_s}\,s_0^2\,\rs
\end{equation}
qualitatively gives signal-to-noise ratio resulting from the Poisson noise 
contribution. 

On the other hand, in estimating the sampling noise, 
a reliable and robust method is the numerical technique 
of bootstrap re-sampling, which has been widely used in the
observational/numerical 
study of genus statistics. Although analytical estimation of 
sampling noise is rather difficult task, we can roughly infer the 
influence of finite sampling effect from {\it number of 
resolution elements} $N_{\rm res}$ which provides a useful 
measure for the statistical power of isodensity statistics 
(Gott et al. 1989; Canavezes et al. 1998).  
{}For a given redshift range of the sample region $[\zmin,\zmax]$, we have 
\begin{equation}
\label{eq: nres}
 N_{\rm res}=\frac{\displaystyle 
   \frac{\Delta\Omega}{3}\,\left\{s^3(\zmax)-s^3(\zmin)\right\}}
  {\displaystyle \frac{4\pi}{3}\rs^3}, 
\end{equation}
where $\Delta\Omega$ denotes a fixed solid angle. Gott et al. (1989) 
introduced this quantity to discuss the quality of data sets and  
argued that for a fractional accuracy of $25\%$ in measuring 
the genus curve, a sample with $N_{\rm res}=100$ is required. 
{}For a measurement with high fractional accuracy of 
$\sim$ few $\%$, large number of resolution element $N_{\rm res}>10^3$
is necessary. Note that in a recent study of Point Source Catalogue 
Redshift Survey of {\it IRAS} galaxies, number of resolution element 
reaches more than $300$ (Canavezes et al. 1998). 
In this paper, we use the quantity $N_{\rm res}$ to estimate the 
influence of the sampling noise.

Table 1 summarizes the signal-to-shot noise ratio (\ref{eq: S/N}) and 
the number of resolution element (\ref{eq: nres}) assuming 
the typical samples of cosmological objects, 
SDSS/2dF quasars, Lyman-break galaxies, and cluster samples. Here, 
we also list the SDSS/2dF galaxies as a representative low-z sample. 
Typical values of the correlation length $s_0$ and the 
mean number density $n_s$ are adopted to evaluate $(S/N)_{\rm shot}$.  
To compute the quantity $N_{\rm res}$, we assume the cosmological 
redshift space $s=\seds(z)$ and set the solid angle $\Delta\Omega$ 
by $\pi$ steradian for the SDSS/2dF galaxies, quasars and clusters, 
$0.25$ square degree for the Lyman-break galaxies. 
From Table 1, it is clear that the SDSS/2dF galaxy redshift surveys 
unambiguously provide us high quality data preferable to an analysis 
of the isodensity statistics. For our interest of high-redshift samples, 
the cluster samples can be good objects for studying the isodensity 
statistics. The systematic cosmological effects can 
dominate the Poisson and the sampling noise, and may act as signal.  
As for the Lyman-break galaxies, while the high signal-to-noise 
ratio $(S/N)_{\rm shot}$ could be attained, the sampling 
error can dominate the signal due to the narrow field of sample region. 
In an optimistic sense, this is not a problem when the wide-field 
samples are supplied. The photometric SDSS galaxies beyond $z\simgt 0.2$ 
might be a candidate of such samples if the redshifts of objects are 
precisely measured, although the redshift range will be shallower $z\simlt 1$.
On the other hand, the quasar samples would be rather sparse and the 
Poisson noise unfortunately dominates the signal, though 
the sampling error becomes negligible.

The above estimation of statistical errors might be rather optimistic 
and idealistic. We do not consider the selection effect and the 
boundary effect of the survey region. Further, we should be aware 
that strength of cosmological effects as well as the statistical 
errors also depend on the biasing of objects, 
which might be one of the most important obstacle in estimating the 
sensitivity. In a rigorous sense, we cannot give any conclusive statements 
until elaborating a more quantitative study using numerical simulation,  
however, we can say that the systematic effects of geometric and velocity 
distortions cannot be entirely negligible. 
The isodensity statistics on clustering of high-redshift 
objects will be measured from the cluster samples, in addition to the 
high-z galaxies from the future wide-field surveys. 
%
%
%
%
%
%
\subsection{Light-cone Effect}
\label{subsec: light-cone}
%
%
%
%
%
%
Table 1 suggests that observations of a 
large volume region is more desirable to 
reduce the sampling noise. In this case, the redshift 
range of the samples becomes large, and the influence of the 
light-cone effect will be important. 
Namely, the time-evolution of the distribution of the
objects should be properly taken into account.
The light-cone effect on the two-point correlation function and 
power spectrum is rigorously formulated by Yamamoto \& Suto (1999) and 
the effect is extensively investigated in various cosmological 
situations (Nishioka \& Yamamoto 1999, 2000; Yamamoto et al. 1999; 
Suto et al. 2000a, b). 

Unlike the two-point statistics, which are computed from the 
pair-count of objects with a fixed separation, 
the isodensity statistics can be estimated from the pixelated 
density fields. This situation 
is similar to the analysis of the higher order moments using 
the count-in-cell method (Matsubara et al. 1997). 
In this case, as long as the distant-observer approximation is 
valid, the light-cone effect is supposed to be incorporated 
into the isodensity statistics as follows: 
  \begin{eqnarray}
    \label{eq: Light-cone}
    G^{\rm \scriptscriptstyle (LC)}_i(\nu) = 
    \frac{\displaystyle \int_{\zmin}^{\zmax} dz 
      \left(\frac{dV_c^{(i)}}{dz}\right)
    G^{(s)}_i(\nu;z) w(z)}{\displaystyle \int_{\zmin}^{\zmax}  dz 
    \left(\frac{dV_c^{(i)}}{dz}\right)w(z)},  
  \end{eqnarray}
where  $G_i^{(s)}(\nu;z)$ denotes the $i$-dimensional genus 
statistics at redshift $z$, $dV_c^{(i)}$ is the comoving 
volume element in $i$-dimensional (cosmological redshift) space,
and $w(z)$ is a weight function. In general, the weight function 
$w(z)$ arising from the selection effect is determined 
so as to maximize the signal-to-noise ratio. As stated 
in Suto et al. (1999), the pixelated density field at $z$ is 
corrected by multiplying the inverse of the selection function 
$\Phi(z)$ for the density fields and $w(z)$ can be set to unity 
for $\zmin<z<\zmax$, in principle. 
This is actually done when we compute the genus curves from the 
observational data (e.g, Canavezes et al. 1998). 
Hence, we simply set $w(z)=1$. 
The formula (\ref{eq: Light-cone}) can also be applicable to the area 
statistics by replacing $G_i^{(s)}(\nu;z)$ with $N_i^{(s)}(\nu;z)$, 
the $i$-dimensional area statistics. 

In the above expression (\ref{eq: Light-cone}), the range 
of the integration $[\zmin, \zmax]$ and the volume factor 
$dV_c^{(i)}$ should be determined depending on geometry 
of the survey depth for the three dimensional statistics, 
and also depending on configuration of a plane (line) underlying
the density fields for the two (one) dimensional statistics. 
As an example, let us consider the case that samples are 
distributed in the range of the radial distance 
$s(\zmin)\leq s \leq s(\zmax)$ within a fixed solid angle 
$\Delta \Omega$.  
In this case, the $i$-dimensional volume factor becomes 
  \begin{equation}
    \label{eq: volumefactor} 
    {dV_c^{(i)}}=  s(z)^{i-1}{ds(z)}\Delta\Omega . 
  \end{equation}

\begin{figure}
\begin{center}
   \leavevmode\epsfxsize=9cm \epsfbox{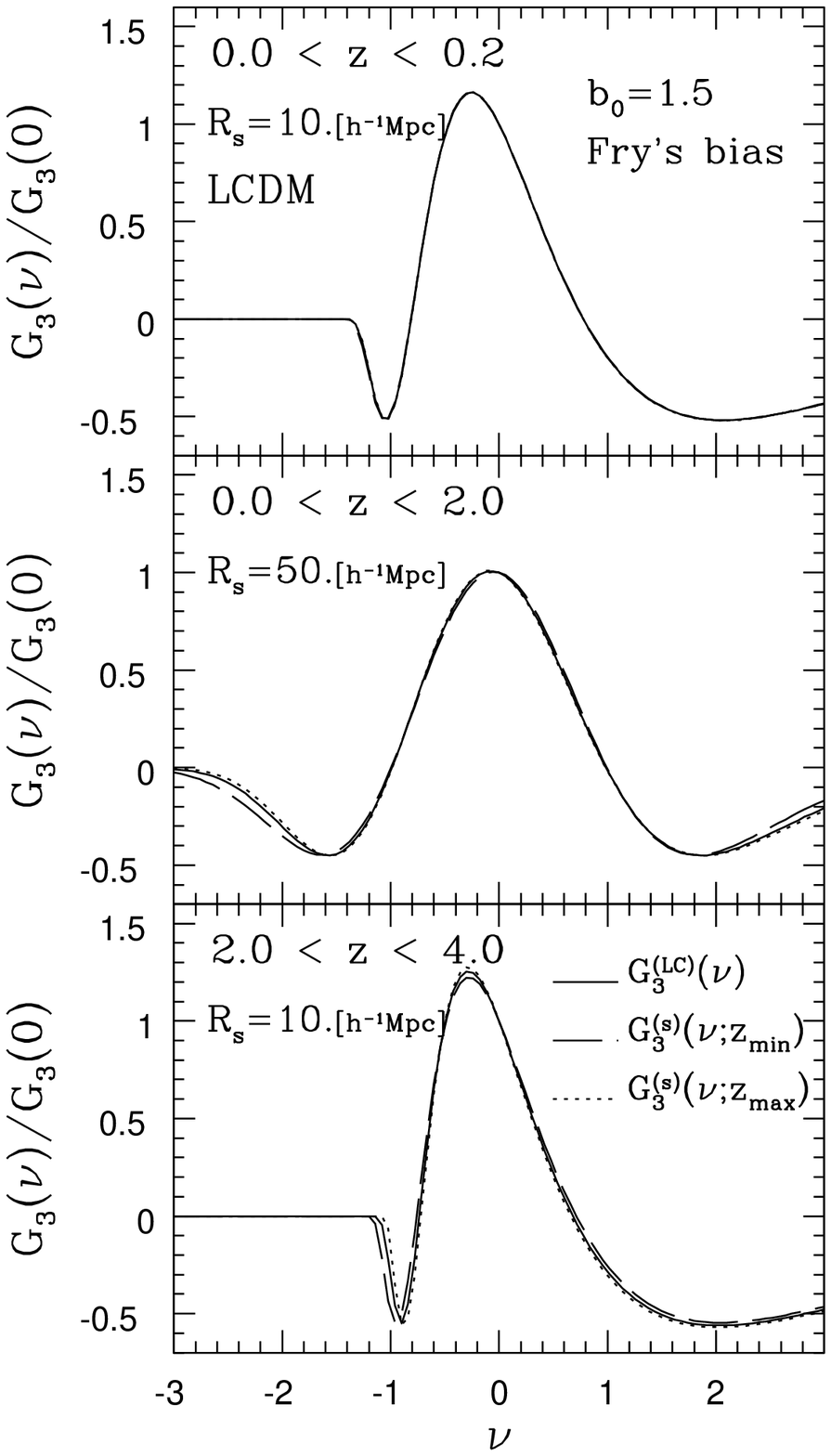}
\hspace*{-2cm}
   \leavevmode\epsfxsize=9cm  \epsfbox{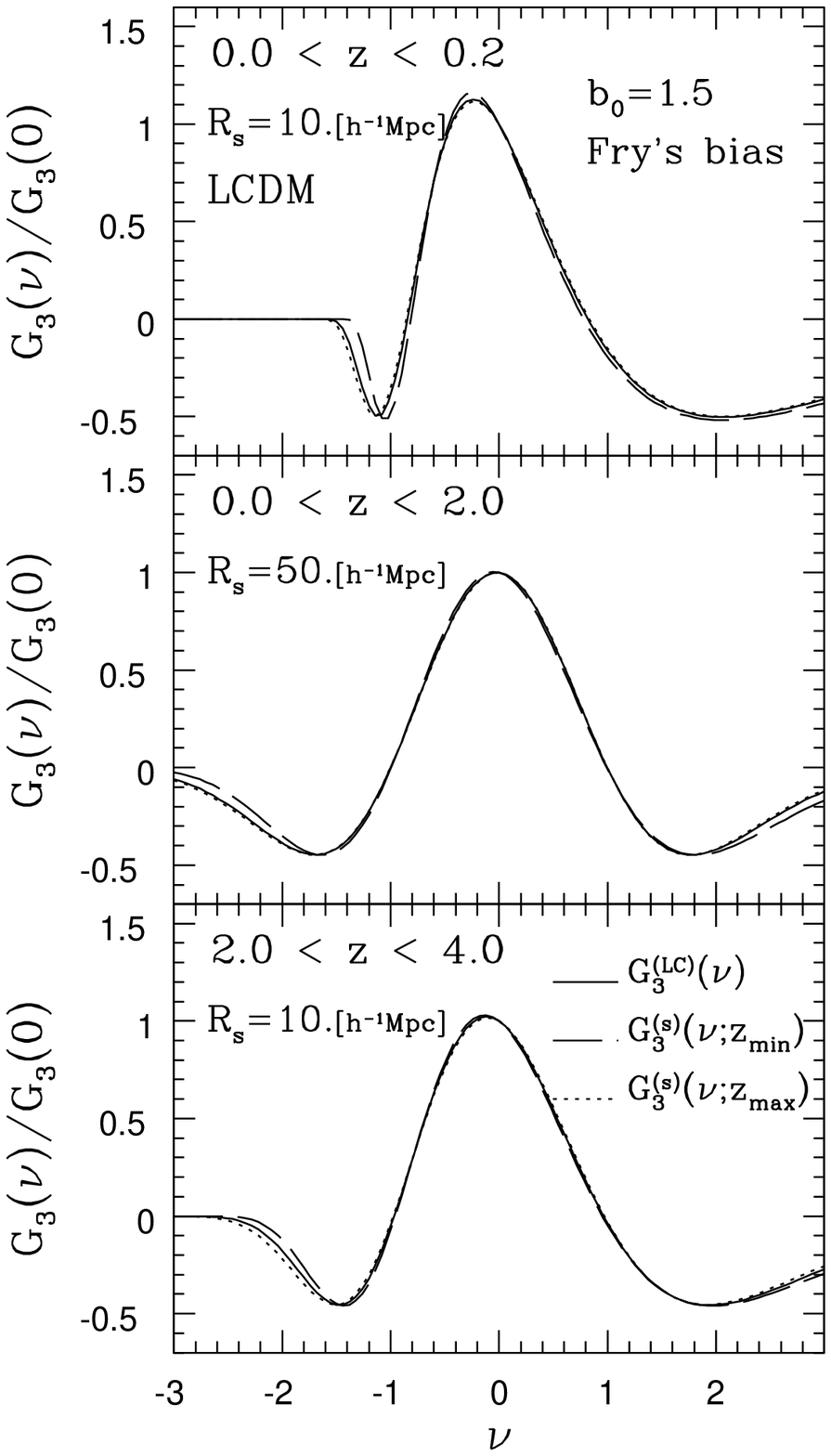}
 \figcaption{The three-dimensional genus curves $G_3$ evaluated on a
 light-cone in the LCDM model. For each panel, the evolution of bias is 
 assumed to be the Fry's model with $b_0=1.5$.
 The cosmological redshift space is chosen as 
 $s=\shb(z)$ in left-panel and $s=\seds(z)$ in right-panel. 
 The adopted range of the redshift-integration and the smoothing
 scale are shown in each panel.
 \label{fig: LCeffect_G3}}

\end{center}
\end{figure}

\begin{figure}
\begin{center}
   \leavevmode\epsfxsize=9cm \epsfbox{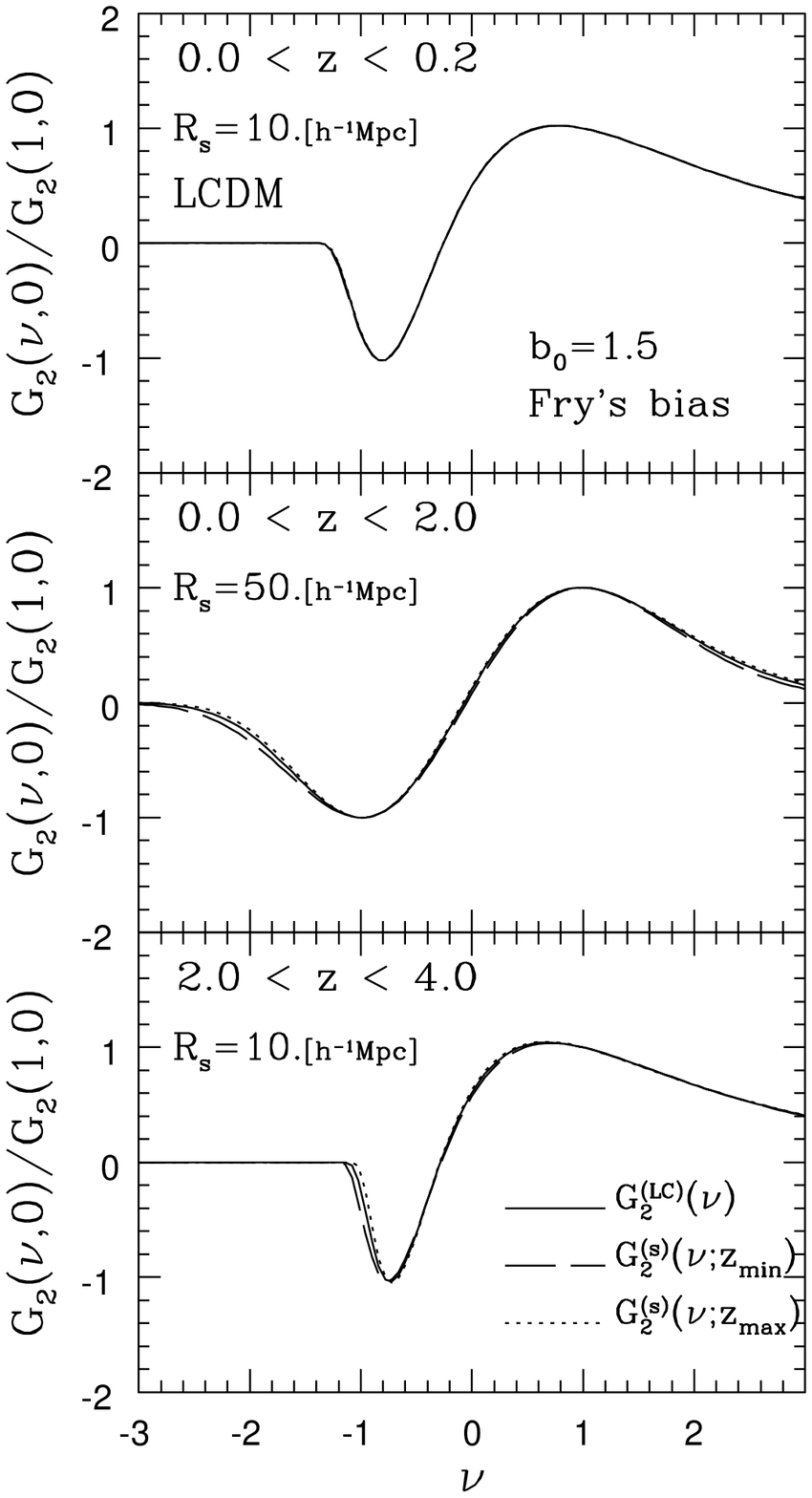}
\hspace*{-2cm}
   \leavevmode\epsfxsize=9cm \epsfbox{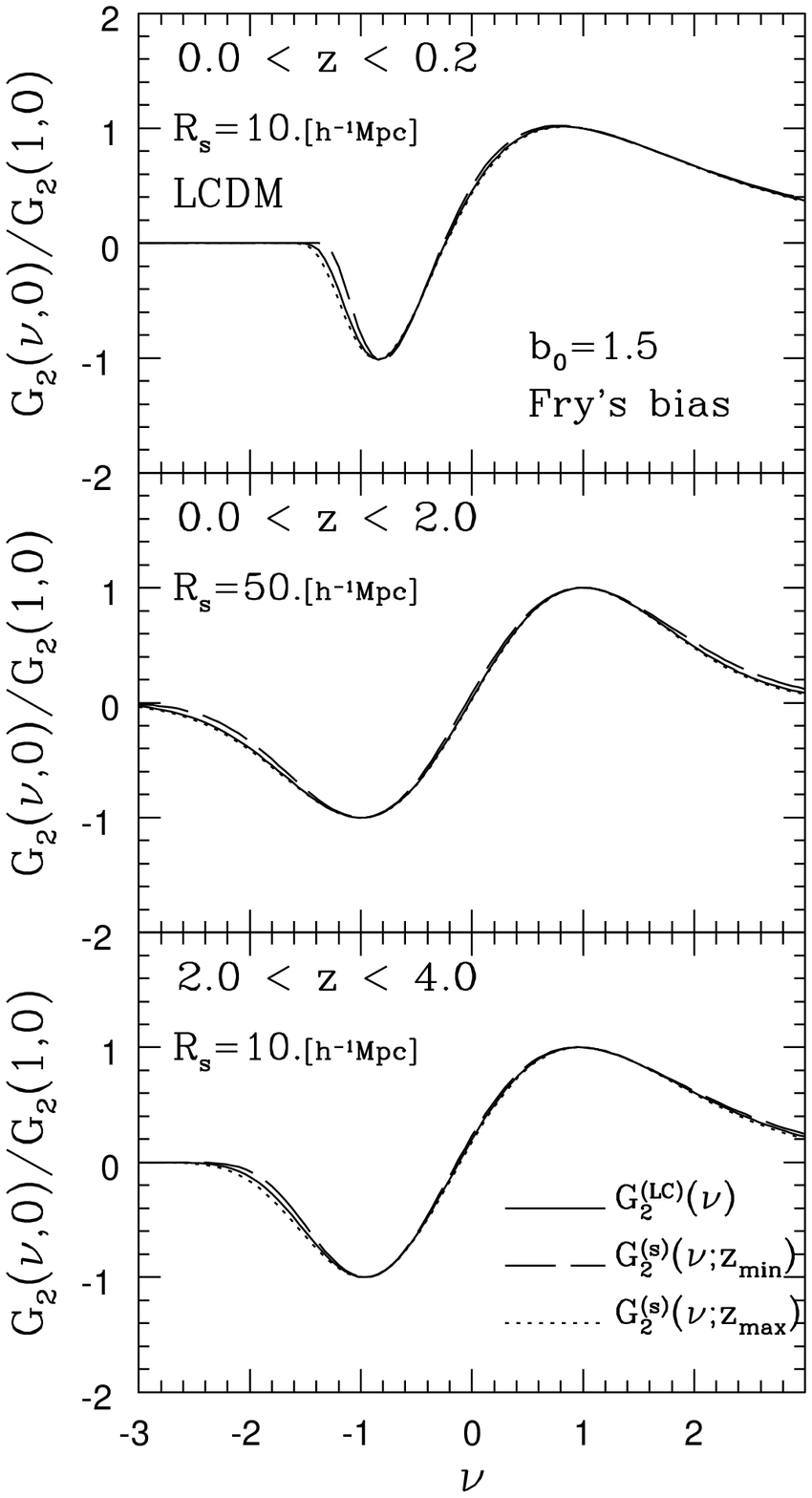}
 \figcaption{Same as Figure \ref{fig: LCeffect_G3}, but the 
   two-dimensional genus curves $G_2$ evaluated on a
 light-cone. Here, we take the two-dimensional slice $S$ 
 parallel to the line-of-sight direction, i.e., $\theta=0$.
\label{fig: LCeffect_G2}}
\end{center}
\end{figure}

As a demonstration, in Figure \ref{fig: LCeffect_G3}, we plot
the three-dimensional genus $G^{\rm \scriptscriptstyle (LC)}_3$ 
for the Lyman-break galaxies ({lower panels}) 
and the high-redshift clusters ({middle panels}) as 
typical high-z samples, and the SDSS/2dF galaxies ({upper panels}) 
as a reference. 
We adopted the parameters of the redshift range of the samples and the
smoothing radius listed in Table 1.
The left and right panels show the cases $s=\shb(z)$ and 
$s=\seds(z)$, respectively. For each panel, solid lines show 
the genus curves incorporating the light-cone effect, 
$G^{\rm \scriptscriptstyle (LC)}_3(\nu)$. On the other hand 
the dashed and dotted lines show the ones on a 
constant hypersurface at $z=\zmin$ and $\zmax$, i.e., 
$G^{\rm \scriptscriptstyle (s)}_3(\nu;\zmin)$ and
$G^{\rm \scriptscriptstyle (s)}_3(\nu;\zmax)$, respectively.
In the similar way, Figure \ref{fig: LCeffect_G2} shows 
the two-dimensional genus $G^{\rm \scriptscriptstyle (LC)}_2$.
In the case of the two-dimensional genus $G_2$, 
the two-dimensional slice $S$ embedded in a survey field 
is taken to be parallel to the line-of-sight direction, i.e., 
$\theta=0$.

From Figures \ref{fig: LCeffect_G3} and \ref{fig: LCeffect_G2},  
we might regard that the light-cone effect is not significant for 
these samples. That is, as long as the samples listed in Table 1 are 
concerned, the light-cone effect on the shapes of isodensity statistics 
is not so significant. This might be because we focused on the shape 
of the isodensity statistics normalized by the amplitude at the 
threshold $\nu=0$ and/or $\nu=1$, which is sensitive to 
probability distribution function of the random field, rather 
than that of the clustering amplitude. 
At a high-redshift and/or with a large smoothing radius, the 
probability distribution as a function of the threshold $\nu$ 
merely has the weak dependence of time.
This consideration suggests that the light-cone effect 
becomes important for smoothing scales on a boundary 
between linear and nonlinear regimes, where the evolution 
of the probability distribution function can be drastic.
%
%
%
%
%
\section{Conclusions}
\label{sec: conclusion}
%
%
%
%
In the present paper, we have considered various observational 
effects on the isodensity statistics of high-redshift objects, 
and estimated their influences, as a generalization of the work 
by Matsubara (1996). 
We have presented a simple theoretical model incorporating the 
geometric distortion and the nonlinear correction arising from 
the gravitational growth and the finger-of-God effects, together with 
useful analytic formulae. From this theoretical model 
(\ref{eq: G_log})--(\ref{eq: N1_log}), we have demonstrated that the 
geometric distortion affects the shapes of the isodensity statistics 
and that the nonlinear correction can be sensitive to a choice 
of the redshift-space coordinate as increasing the redshift. 
Based on these results, we have briefly discussed the detectability 
of these effects by evaluating the shot-noise and 
the sampling noise contaminations. Furthermore,  
the influence of the light-cone effect has been investigated.  
The important conclusions are summarized:  
\begin{enumerate}
\item 
When the cosmological redshift space $s=\shb(z)$ 
(eqs.[\ref{eq: shb}] and [\ref{eq: z_crd}]) is chosen, the nonlinear 
correction has an effect on the isodensity statistics for smoothing 
length larger $\rs\simeq 10h^{-1}$Mpc even at the higher 
redshift $z\simgt1$, where the isodensity statistics have 
asymmetric shapes. On the other hand, when the cosmological 
redshift space $s=\seds(z)$ (eqs.[\ref{eq: seds}] and [\ref{eq: EdS_crd}]) 
is chosen, the nonlinear correction turns out to be small and the 
linear theory of redshift-space distortions using the 
Gaussian results (\ref{eq: G3_Gauss}), (\ref{eq: G2_Gauss}) 
and (\ref{eq: N3_Gauss})-(\ref{eq: N1_Gauss}) could be a 
good approximation for the large smoothing scales. 
In both cases, the geometric and velocity distortions cause 
the angle-dependence in the low-dimensional 
quantities, and the relative amplitude can be changed by $10\sim20\%$.  

\item 
The Poisson sampling error is crucial for the sparse sampling 
of high-redshift objects like the (SDSS/2dF) quasars. 
The finiteness of the sampling volume might affect the precise 
measurement of isodensity statistics with Lyman-break galaxies. 
Though more quantitative estimation of statistical errors 
using numerical simulations is needed, however,
the isodensity statistics can be tested using clusters 
and high-z galaxies from future wide-field surveys.  
For these samples, the cosmological effects 
can be comparable or could be dominant to the Poisson and the sampling noise. 

\item 
For a sample that the survey volume 
is large and the cosmological objects are distributed in a wide
range of redshift, the light-cone effect can be influential.
However, we infer that the light-cone effect has a weak influence 
on the isodensity statistics as long as the time evolution of 
the probability distribution function is not drastic, though 
this picture might depend on the evolution of the biasing factor.
\end{enumerate}
Because the model presented here 
are quite simple, the predictions should be checked 
by numerical simulations. The prediction can also be 
sophisticated by using more reliable nonlinear models 
deduced from the simulations. Recent development of cosmological 
N-body simulation enables us to obtain the enormous large 
volume data over a few Gpc (Colberg et al. 2000). The larger 
data comparable to the Hubble volume size will provide us with 
chances of quantitative studies for the isodensity statistics,
as well as the two-point statistics, in which the light-cone 
effect inevitably becomes important.

Combining our theoretical template with these simulations, various 
cosmological implications might be yielded. One of the most interesting 
application is the geometric test proposed by Alcock \& Paczy\'nski 
(1979). As we have seen in Sec.\ref{subsec: results}, the geometric 
distortion makes the observed structures distorted and it causes 
deformation of the isodensity contour. 
Since this effect is sensitive 
to the distance-redshift relation, the anisotropy of isodensity 
statistics might be a useful tool in investigating cosmological
models, as well as the anisotropy of two-point statistics 
(Yamamoto, Nishioka \& Taruya 2000). 
The present theoretical formulae include 
useful tools for quantifying the geometric distortion, as well as for testing 
statistical nature of biased density fields. 
To address these issues, however, a more quantitative analysis should 
be developed.

The uncertainty of the biasing can be a critical issue to clarify 
the topological features of large-scale structure. Throughout the analyses, 
we have put the linear biasing ansatz (\ref{eq: bias}) 
and the evolution of the biasing simply given by the model 
(\ref{eq: FryBias}) neglecting the merging and formation processes. 
In more realistic situation, the biasing of luminous objects is 
nonlinear and stochastic (Dekel \& Lahav 1999) and the merging 
and formation processes might alter the evolution of biasing 
(e.g, Somerville et al. 2000; Taruya et al. 2000). 
However, recent theoretical and numerical studies suggest 
that the linear biasing assumption (\ref{eq: bias}) can be a 
good approximation even in the quasi-linear regime 
(Coles, 1993; Scherrer \& Weinberg 1998; 
Matsubara 1999; Taruya \& Suto 2000; 
Taruya et al. 2000). Therefore, in a qualitative 
sense, we believe that the present analyses would have revealed 
the generic feature of isodensity statistics on the observed 
high-redshift objects. As for the evolution of biasing, the 
theoretical prediction based on the Press-Schechter formalism 
gives the biasing factor for the dark halo distribution 
accurately (Mo \& White 1996; Jing 1998; Sheth \& Tormen 1999). 
Incorporating this evolution model into our analytic formulae 
can provide a powerful tool for quantifying the distribution of 
clusters of galaxies. 
In any way the isodensity statistics 
may probe a distinctive aspect of the biasing factor. 
This will be discussed elsewhere (Hikage, Taruya \& Suto 2000, 
in preparation).

\bigskip
\bigskip

We thank Prof. Suto for useful discussion and 
critical comments, Prof. Matsubara for useful discussion. 
A.T. gratefully acknowledges support from a JSPS 
(Japan Society for the Promotion of Science) fellowship.  K.Y. thanks Prof. Kojima and Mr. 
Nishioka for useful comments. This work is supported by the 
Inamori foundation and in part by the Grants-in-Aid program (11640280) 
by the Ministry of Education, Science, Sports and Culture of Japan.  
\clearpage
%
%
%

%
%
%
%
%
%
%
%
\clearpage
\appendix
%
%
%
%
\section{Analytic formulae for Gaussian random field}
\label{appendix: gaussian results}
%
%
%
%
In this appendix, we give the analytic formulae of 
isodensity statistics, which have been previously 
derived by Matsubara (1996). Although he only consider the case  
in absence of the geometric distortion (i.e, $\cpara=\cperp=1$) 
by imposing the assumption of linear redshift-space distortion,   
his results can be easily extended to those incorporating 
the geometric distortion in a rather formal way, 
without any approximation except for the random Gaussianity. 

Let us suppose that the density field is homogeneous random Gaussian.   
In this case, the joint probability 
distribution function ${\cal P}(\deltas,\,\eta_i,\,\zeta_{ij})$ is 
divided into ${\cal P}(\deltas,\,\zeta_{ij}){\cal P}(\eta_i)$ 
regardless of the anisotropy. 
Final expression for the three dimensional genus 
(\ref{eq: def_Genus}) is immediately obtained by following similar 
calculations in the homogeneous and isotropic case 
(Bardeen et al. 1986; Hamilton et al. 1986). 
The resultant formula does not 
alter the shape of the genus as a function of $\nu$. Only the amplitude 
is changed\footnote{This has been partially shown by Matsubara (1996) 
in the case of the linear redshift distortion.}:  
\begin{equation}
\label{eq: G3_Gauss}
  G_3^{(s)}(\nu)=\frac{1}{4\pi^2}
  \frac{\sigmaperp^2\sigmapara}{\sigmavar^3}\,\,
   (1-\nu^2)\,\,e^{-\nu^2/2}. 
\end{equation}

On the other hand, for the two-dimensional genus 
(\ref{eq: def_2Dgenus}), by using the relations
\begin{eqnarray}
&& \left\langle\delta_D(\deltas-\nu\sigmavar)
\delta_D(\eta_{\scriptscriptstyle 1})
\,\,|\eta_{\scriptscriptstyle 2}\sin{\theta}
+\eta_{\scriptscriptstyle 3}\cos{\theta}|\,\,
\zeta_{\scriptscriptstyle 11} \right\rangle 
\nonumber 
\\ 
&&~~~~~~~~~~~~~~~~~~~~~~~~~~~~~~~~~=
 \left\langle\delta_D(\deltas-\nu\sigmavar)\,\zeta_{\scriptscriptstyle 11}
\right\rangle\,\, \left\langle 
  \delta_D(\eta_{\scriptscriptstyle 1})\,\,
\left|\eta_{\scriptscriptstyle 2}\sin{\theta}
+\eta_{\scriptscriptstyle 3}\cos{\theta}\right|\,\,
 \right\rangle,  
\nonumber
\\
&&\left\langle\delta_D(\deltas-\nu\sigmavar)\,\zeta_{\scriptscriptstyle 11}
\right\rangle =
\frac{-1}{\sqrt{8\pi}}\left(\frac{\sigmaperp}{\sigmavar}\right)^2
\nu\,e^{-\nu^2/2},
\nonumber
\end{eqnarray}
and 
\begin{eqnarray}
&\left\langle 
  \delta_D(\eta_{\scriptscriptstyle 1})\,
\left|\eta_{\scriptscriptstyle 2}\sin{\theta}
+\eta_{\scriptscriptstyle 3}\cos{\theta}\right|\,
 \right\rangle &=\,\,\, 
\frac{1}{\sqrt{2\pi}}\,L(\gamma,\,\theta),
\nonumber 
\end{eqnarray}
one obtains 
\begin{equation}
  G_2^{(s)}(\nu,\theta) = 
\frac{1}{8\pi}
  \left(\frac{\sigmaperp}{\sigmavar}\right)^2\,\,
   \nu\,\,e^{-\nu^2/2}\,\,L(\gamma,\theta), 
\label{eq: G2_Gauss}
\end{equation}
where the function $L$ is defined by (\ref{app_B: L}) in 
Appendix \ref{appendix: result_JKL}. In the above expression, 
we defined the distortion parameter $\gamma$ given by 
(see also eq.[\ref{eq: gamma}]) 
\begin{eqnarray}
\nonumber
      \gamma \equiv \sqrt{2}\,\frac{\sigmapara}{\sigmaperp}.    
\end{eqnarray}

Similar to the case of the genus statistics, the area statistics, $N_3$, 
$N_2$, and $N_1$ are calculated using the probability functions being 
${\cal P}(\deltas,\,\eta_i)={\cal P}(\deltas){\cal P}(\eta_i)$, which 
is entirely characterized by the field correlations (\ref{eq: var1}) 
and (\ref{eq: var2}). After some manipulations, one can obtain the 
final expressions: 
\begin{eqnarray}
\label{eq: N3_Gauss}
& N_3^{(s)}(\nu) \,=&\, \frac{1}{2\sqrt{\pi}}\frac{\sigmaperp}{\sigmavar}
  ~e^{-\nu^2/2}~~J(\gamma),
\\
\label{eq: N2_Gauss}
& N_2^{(s)}(\nu,\theta) \,=&\, \frac{1}{2\sqrt{\pi}}\frac{\sigmaperp}{\sigmavar}
  ~e^{-\nu^2/2}~~K(\gamma,\theta),
\\
\label{eq: N1_Gauss}
& N_1^{(s)}(\nu,\theta) \,=&\, \frac{1}{2\sqrt{\pi}}\frac{\sigmaperp}{\sigmavar}
  ~e^{-\nu^2/2}~~L(\gamma,\theta),  
\end{eqnarray}
where $L$ is the same function as defined in (\ref{eq: G2_Gauss})  
and the functions $J$ and $K$ are also defined in Appendix B by 
(\ref{app_B: J}) and (\ref{app_B: K}), respectively.
%
%
%
%
%
%
\section{Expressions of $J(\gamma)$, $K(\gamma)$ and $L(\gamma)$}
\label{appendix: result_JKL}
%
%
%
%
%
We here present explicit formulae for the functions  
$J$, $K$ and $L$ to describe the isodensity statistics for 
Gaussian random field (Appendix \ref{appendix: gaussian results}).
The functions $J$, $K$ and $L$ are defined by the ensemble 
average of the Gaussian random field $\eta$ as follows: 
%
%
\begin{eqnarray}
& J(\gamma)\equiv &\frac{\sqrt{2}}{\sigmaperp}\,\,\,
  \langle\,\,\sqrt{\eta_1^2+\eta_2^2+\eta_3^2}\,\,\rangle,
\nonumber 
\\
& K(\gamma,\theta)\equiv &\frac{\sqrt{2}}{\sigmaperp}\,\,\,
  \langle\,\,\sqrt{\eta_1^2+(\eta_2\sin{\theta}+\eta_3\cos{\theta})^2}\,\,\rangle,
\nonumber 
\\
& L(\gamma,\theta)\equiv &\frac{\sqrt{2}}{\sigmaperp}\,\,\,
  \langle\,\, \left|\eta_2\sin{\theta}+\eta_3\cos{\theta}\right|\,\,\rangle, 
\nonumber
\end{eqnarray}
where the ensemble average $\langle\cdots\rangle$ is evaluated 
according to the Gaussian probability distribution function 
\begin{eqnarray}
  {\cal P}(\eta_i)d\eta_1d\eta_2d\eta_3
\,\,=\,\,\frac{1}{(2\pi)^{3/2}(\sigmapara\sigmaperp^2/2)}\,\,\,
   \exp\left[-\frac{\eta_1^2+\eta_2^2}{\sigmaperp^2}
        -\frac{\eta_3^2}{2\sigmapara^2}\right]\,\,d\eta_1d\eta_2d\eta_3.
\nonumber
\end{eqnarray}
{}From straightforward calculations, we obtain the following formulae. 
Note that these formulae are also used to obtain the results 
for the parameterized non-Gaussian case if we 
replace the quantities $\eta_i$, $\sigmapara^2$ and $\sigmaperp^2$ 
with $\partial_i\phi$, $\langle\, [\partial_3\phi]^2\, \rangle_{\phi}$ 
and $\langle\, [\partial_{\bot}\phi]^2\, \rangle_{\phi}$, respectively. 
(see Section \ref{subsec: modeling} and Appendix 
\ref{appendix: nonGaussian}).  
%
%
%
\subsection{$J(\gamma)$}
%
\begin{eqnarray}
 J(\gamma)&=&\int\frac{dxdydz}{(2\pi)^{3/2}}~~e^{-(x^2+y^2+z^2)/2}~
  \sqrt{x^2+y^2+\gamma^2z^2}
\nonumber\\
 &=&\sqrt{\frac{2}{\pi}}\frac{Q^2}{\gamma}~~~\times
\left\{
\begin{array}{lr}
-\frac{\displaystyle 1}{\displaystyle Q(1-Q)}-
        \frac{\displaystyle 1}{\displaystyle 2Q\sqrt{Q}}
 \ln\left|\frac{\displaystyle 1-\sqrt{Q}}{\displaystyle 1+\sqrt{Q}}\right| 
        &;~~~(|\gamma|>1)
\\
\\
-\frac{\displaystyle 1}{\displaystyle Q(1-Q)}-
        \frac{\displaystyle 1}{\displaystyle Q\sqrt{|Q|}}
 \tan^{-1}\left(\frac{\displaystyle 1}{\displaystyle \sqrt{|Q|}}\right) 
        &;~~~(|\gamma|<1)
\end{array}
\right.,
\label{app_B: J}
\end{eqnarray}
where we define $Q=\gamma^2/(\gamma^2-1)$.
%
%
%
%
\subsection{$K(\gamma,\theta)$}
%
\begin{eqnarray}
 K(\gamma,\theta)&=&\int\frac{dxdydz}{(2\pi)^{3/2}}~~e^{-(x^2+y^2+z^2)/2}~
  \sqrt{x^2+(y\sin{\theta}+\gamma~z\cos{\theta})^2},
\nonumber \\
 &=& 
\left\{
\begin{array}{lr}
\sqrt{\frac{2}{\pi}}~\sqrt{1+(\gamma^2-1)\cos^2{\theta}}~
  E\left(\sqrt{\frac{\displaystyle (\gamma^2-1)\cos^2{\theta}}
    {\displaystyle 1+(\gamma^2-1)\cos^2{\theta}}}\right)
  &;~~(|\gamma|>1)
\\
\\
\sqrt{\frac{2}{\pi}}~
  E\left(\sqrt{(1-\gamma^2)\cos^2{\theta}}\right)  &;~~(|\gamma|<1)
\end{array}
\right., 
\label{app_B: K}
\end{eqnarray}
where $E(k)$ is the complete elliptical integral of the second kind, 
\begin{equation}
 E(k)=\int_{0}^{\pi/2}d\phi\,\,\sqrt{1-k^2\sin^2{\phi}}.
\end{equation}
%
%
%
%
\subsection{$L(\gamma,\theta)$}
%
%
\begin{eqnarray}
 L(\gamma,\theta)&=&\int\frac{dxdydz}{(2\pi)^{3/2}}~~e^{-(x^2+y^2+z^2)/2}~
  \left|~y\sin{\theta}+\gamma~z\cos{\theta}~\right|
\nonumber \\
&=& \sqrt{\frac{2}{\pi}}\left|~1+(\gamma^2-1)\cos^2{\theta}~\right|^{1/2}.
\label{app_B: L}
\end{eqnarray}
%
%
%
%
%
%
\section{Analytic results for the parameterized non-Gaussian case}
\label{appendix: nonGaussian}
%
%
%
%
In this appendix, we derive analytic expressions for the isodensity 
statistics with a non-Gaussian distribution whose statistical 
property is characterized by a monotonic function of the random 
field $\phi(\s)$ (see eq.[\ref{eq: del-phi}]). 
Since the field $\phi(\s)$ is a Gaussian variable with the zero mean 
and the unit variance, the one-point distribution function of the 
density fluctuations $P(\deltas)$ becomes 
\begin{eqnarray}
 {\cal P}(\deltas)\,\,d\deltas\,\,&=&\,\, 
\frac{1}{\sqrt{2\pi}}\,\frac{\exp\left\{-\frac{1}{2}[F^{-1}(\deltas)]^2\right\}}
{\left|F_{,\phi}[F^{-1}(\deltas)]\right|}\,\,d\deltas  
\nonumber \\
&=&\,\, \frac{1}{\sqrt{2\pi}}\,\,e^{-\phi^2/2}\, d\phi.
\label{C: PDF_deltas}
\end{eqnarray}
Recall that the ensemble average with respect to the density field 
$\deltas$ can be computed through the average with 
respect to the Gaussian field $\phi$. Using the fact that 
the field variables $\eta_i$ and $\zeta_{ij}$ can 
be written by 
\begin{eqnarray}
 &&\eta_i=F_{,\phi}\partial_i\phi,
\,\,\,\,\,\,\,\,\,\,
\nonumber
\\
 &&\zeta_{ij}=F_{,\phi}\,\partial_i\partial_j\phi +F_{,\phi\phi}\,
\partial_i\phi\,\partial_j\phi,
\nonumber
\end{eqnarray}
the final results are obtained : 
\begin{eqnarray}
 G_3^{(s)}(\nu)\,&=& 
-\frac{1}{2}\,\left\langle\,\delta_D[F(\phi)-\nu\sigmavar]\,\,
\delta_D[F_{,\phi}\partial_1\phi]\,\,\delta_D[F_{,\phi}\partial_2\phi]
\,\,|F_{,\phi}\partial_3\phi|\right.
\nonumber \\
&\times& \,
\left.\left\{[F_{,\phi}\,\partial_1\partial_1\phi +F_{,\phi\phi}\,
(\partial_1\phi)^2]
[F_{,\phi}\,\partial_2\partial_2\phi +F_{,\phi\phi}\,
(\partial_2\phi)^2]
-[F_{,\phi}\,\partial_1\partial_2\phi +F_{,\phi\phi}\,
\partial_1\phi\,\partial_2\phi]^2
\right\} \,\right\rangle_{\phi} 
\nonumber \\
&=& 
-\frac{1}{2}\,\left\langle\,\delta_D[\phi-F^{-1}(\nu\sigmavar)]\,\,
\delta_D[\partial_1\phi]\,\,\delta_D[\partial_2\phi]\,\,|\partial_3\phi|\,\,
\left\{(\partial_1\partial_1\phi)(\partial_2\partial_2\phi)
-(\partial_1\partial_2\phi)^2\right\} \,\right\rangle_{\phi} 
\nonumber \\
&=&\frac{1}{(2\pi)^2}\,\,
\frac{\sigmaperp^2\sigmapara}
  {\left\{\langle F_{,\phi}(\phi)^2\rangle_{\phi}\right\}^{3/2}}\,\,
\,\left\{1-[F^{-1}(\nu \sigmavar)]^2\right\}
\exp\left\{-\frac{1}{2}\left[F^{-1}(\nu\sigmavar)\right]^2\right\} .
\label{C: G_phi}
\end{eqnarray}
for the three-dimensional genus (see also Matsubara \& Yokoyama 1996), 
where the ensemble average 
$\langle\cdots\rangle_{\phi}$ is taken with respect to the field $\phi$. 
{}For the two-dimensional genus, we have
\begin{eqnarray}
 G_{2}^{(s)}(\nu,\theta)\,&=&\,
-\frac{1}{2}\,\left\langle\,\delta_D[F(\phi)-\nu\sigmavar]\,\,
\delta_D[F_{,\phi}\partial_1\phi]\,\,
\left|F_{,\phi}(\partial_2\phi\,\sin{\theta}+\partial_3\phi\,\cos{\theta})
\right| \right.
\nonumber \\
&&~~~~~~~~~~~~~~~~~~~~~~~~~~~~~~~~~~~~~~~~~~~~~~~~~~~~~~~~\times \,\left.
[F_{,\phi}\,\partial_1\partial_1\phi +F_{,\phi\phi}\,
(\partial_1\phi)^2]\,\,\right\rangle_{\phi} 
\nonumber \\
\,&=&\,
-\frac{1}{2}\,\left\langle\,\delta_D[\phi-F^{-1}(\nu\sigmavar)]\,\,
\delta_D[\partial_1\phi]\,\,
\left|\partial_2\phi\,\sin{\theta}+\partial_3\phi\,\cos{\theta}\right|
\,\,\partial_{1}\partial_{1}\phi\,
\,\,
\right\rangle_{\phi} 
\nonumber \\
&=&\,\,\frac{1}{8\pi}\,\,
\frac{\sigmaperp^2}
  {\langle F_{,\phi}(\phi)^2\rangle}_{\phi}\,\,
F^{-1}(\nu \sigmavar)
\exp\left\{-\frac{1}{2}\left[F^{-1}(\nu\sigmavar)\right]^2\right\}\,\,
  L(\gamma) ,
\label{C: G2_phi}
\end{eqnarray}
where $\gamma$ is given by (\ref{eq: gamma}) and 
we have used the formula (\ref{app_B: L}) when taking the 
ensemble average with respect to the Gaussian field $\partial_i\phi$. 
As for the area statistics $N_3$, $N_2$ and
$N_1$, repeating the similar calculations yields  
\begin{equation}
 N_{3}^{(s)}(\nu)= \frac{1}{2\sqrt{\pi}}\,\,
\frac{\sigmaperp}
  {\sqrt{ \langle F_{,\phi}(\phi)^2\rangle}_{\phi}}\,\,
\exp\left\{-\frac{1}{2}\left[F^{-1}(\nu\sigmavar)\right]^2\right\}\,\,
  J(\gamma),
\label{C: N3_phi}
\end{equation}
\begin{equation}
 N_{2}^{(s)}(\nu,\theta)=
\frac{1}{2\sqrt{\pi}}\,\,
\frac{\sigmaperp}
  {\sqrt{ \langle F_{,\phi}(\phi)^2\rangle}_{\phi}}\,\,
\exp\left\{-\frac{1}{2}\left[F^{-1}(\nu\sigmavar)\right]^2\right\}\,\,
  K(\gamma),
\label{C: N2_phi}
\end{equation}
\begin{equation}
 N_{1}^{(s)}(\nu,\theta)=
\frac{1}{2\sqrt{\pi}}\,\,
\frac{\sigmaperp}
  {\sqrt{ \langle F_{,\phi}(\phi)^2\rangle}_{\phi}}\,\,
\exp\left\{-\frac{1}{2}\left[F^{-1}(\nu\sigmavar)\right]^2\right\}\,\,
  L(\gamma).
\label{C: N1_phi}
\end{equation}
In deriving the above formulae, we have used the following relations,
\begin{eqnarray}
& \sigmavar^2 =& \langle F(\phi)^2\rangle_{\phi}\,\,, 
\nonumber 
\\
 & \sigmapara^2 =& \langle F_{,\phi}(\phi)^2\rangle_{\phi}\,
  \langle\, [\partial_3\phi]^2\, \rangle_{\phi}\,\,, 
\nonumber 
\end{eqnarray}
and
\begin{equation}
 \frac{\sigmaperp^2}{2} = \langle F_{,\phi}(\phi)^2\rangle_{\phi}\,
  \langle\, [\partial_J\phi]^2\,\rangle_{\phi}\,\,,  
\nonumber
\end{equation}
where the subscript $_J$ runs from 1 to 2.
%
%
%
%
%
%
%
%
%
%
%
%
%
%
%
\begin{deluxetable}{lcccccccc}
\tablecolumns{10}
\tablecaption{
  High-redshift samples and signal-to-noise ratios of 
  isodensity statistics. As a representative low-z sample, 
  we also list the SDSS/2dF 
 galaxy samples. Typical values from observations 
  are adopted for the correlation length $s_0$ and the mean 
  number density $n_s$. The smoothing length $\rs$ is 
  specifically chosen as $\rs=2.5 s_0$ except for the quasar samples,  
  $\rs=50 h^{-1}$Mpc. Then signal-to-noise ratio 
  $(S/N)_{\rm shot}$ is evaluated from definition (\ref{eq: S/N}). 
  Number of resolution element $N_{\rm res}$ is computed from 
  (\ref{eq: nres}) assuming the cosmological redshift space $s=\seds(z)$ 
 and setting the solid angle of survey field $\Delta\Omega$ by 
 $\pi$ steradian (SDSS galaxies, quasars and clusters) or $0.25$ square 
 degree (Lyman-break galaxies). 
}
\tablewidth{17cm}
\tablehead{
 \colhead{samples} & $[\zmin,\zmax]$ & 
 \colhead{$s_0[h^{-1}$Mpc]} & \colhead{$\rs[h^{-1}$Mpc]} & 
 \colhead{$n_s[h^{3}$Mpc$^{-3}]$} & 
 \colhead{$(S/N)_{\rm shot}$} & 
 \colhead{$N_{\rm res}$} 
}
\startdata
SDSS/2dF galaxies & $[0.0,~0.2]$ &  4.0 \phs & 10 \phs 
&  0.1 \phs & 710 \phs & $3.6 \times 10^4$\phs \\
Clusters & $[0.0,~2.0]$ &  20.0 \phs & 50 \phs &  
$10^{-5}$ \phs & 9.0 \phs & $3.3\times10^4$\phs \\
Lyman-break galaxies & $[2.0,~4.0]$ &  4.0 \phs & 10 \phs 
& $10^{-2}$ \phs  & 360 \phs  &  $1.2\times10^2$\phs \\
SDSS/2dF quasars & $[0.0,~3.0]$ & 4.0 \phs & 50 \phs 
& $10^{-6}$ \phs & 0.04 \phs & $5.4\times10^4$\phs\\
\enddata
\end{deluxetable}
\clearpage
%
%
%
%
%
%
%
%
%
%
%
%
%
%
%
%
%
%
%
%
%
%
%
%
%
%
%
%
%
%
%
%
%
%
%
%
%
%
%
%
%
%
%
%
%
%
%
%
%
%
%
%
\newpage
\vspace{0.5cm}
%
%
%
%
%
%
%
%
%
\end{document}